\documentclass[12pt,preprint]{aastex}

\newcommand{\etal}{et~al.\ }

\newcommand{\feka}{\hbox{Fe\,K$\alpha$}}

\newcommand{\cmsq}{\hbox{cm$^{-2}$}}

\newcommand{\simgt}{\lower 2pt \hbox{$\, \buildrel {\scriptstyle >}\over {\scriptstyle\sim}\,$}}
\newcommand{\simlt}{\lower 2pt \hbox{$\, \buildrel {\scriptstyle <}\over {\scriptstyle\sim}\,$}}

\newcommand{\nh}{\hbox{${N}_{\rm H}$}}
\newcommand{\aox}{$\alpha_{\rm ox}$}

\newcommand{\eg}{e.g.,\,}
\newcommand{\ie}{i.e.,\,}
\newcommand{\be}{\begin{equation}}
\newcommand{\ee}{\end{equation}}
\newcommand{\bea}{\begin{eqnarray}}
\newcommand{\eea}{\end{eqnarray}}
\newcommand{\nhat}{\hat{\bf n}}

\shorttitle{\emph{CHANDRA} OBSERVATIONS of GRAVITATIONAL LENSES}
\shortauthors{CHEN ET AL.}

\tighten

\begin{document}

%--------------------------------------------------------------------------------------
\def\sarc{$^{\prime\prime}\!\!.$}
\def\arcsec{$^{\prime\prime}$}
\def\arcmin{$^{\prime}$}
\def\degr{$^{\circ}$}
\def\seco{$^{\rm s}\!\!.$}
\def\ls{\lower 2pt \hbox{$\;\scriptscriptstyle \buildrel<\over\sim\;$}}
\def\gs{\lower 2pt \hbox{$\;\scriptscriptstyle \buildrel>\over\sim\;$}}

\title{Inclination-Dependent AGN Flux Profiles From Strong Lensing of the Kerr Space-Time }

\author{Bin Chen\altaffilmark{1}, Xinyu Dai\altaffilmark{1}, E. Baron\altaffilmark{1,2,3,4} }

\altaffiltext{1}{Homer L. Dodge Department of Physics and Astronomy, The University of Oklahoma,
Norman, OK, 73019, USA, bchen@ou.edu}
\altaffiltext{2}{Computational Research Division, Lawrence Berkeley
        National Laboratory, MS 50F-1650, 1 Cyclotron Rd, Berkeley, CA}
\altaffiltext{3}{ Physics Department, Univ. of California,  Berkeley, CA 94720 USA }
\altaffiltext{4}{Hamburger Sternwarte, Gojenbergsweg 112, 21029 Hamburg, Germany }

\begin{abstract}
Recent quasar microlensing observations have constrained the X-ray emission sizes of quasars to be about $10$ gravitational radii, one order of magnitude smaller than the optical emission sizes. 
Using a new ray-tracing code for the Kerr space-time, we find that the observed X-ray flux is strongly influenced by the gravity field of the central black hole,  even for observers at moderate inclination angles. 
We calculate inclination-dependent flux profiles of active galactic nuclei in the optical and X-ray bands by combining the Kerr lensing and projection effects for future references.
We further study the dependence of the X-ray-to-optical flux ratio on the inclination angle caused by differential lensing distortion of the X-ray and optical emission, assuming several corona geometries.
The strong lensing  X-ray-to-optical magnification ratio  can change by a factor of $\sim$10 for normal quasars in some cases, and another factor of $\sim$10 for broad absorption line quasars (BALs) and obscured quasars.
Comparing our results with the observed distributions in normal and broad absorption line quasars, we find that the inclination angle dependence of the magnification ratios can change the X-ray-to-optical flux ratio distributions significantly.
In particular, the mean value of the spectrum slope parameter $\alpha_{\rm ox},$  $0.3838\log F_{\rm 2\, keV}/F_{\rm 2500\, {\scriptsize \AA}}$, can differ by $\sim$0.1--0.2 between normal and broad absorption line quasars, depending on corona geometries, suggesting larger intrinsic absorptions in BALs.   
\end{abstract}


%-----------------------------------------
\section{Introduction}

X-ray emission is a defining characteristic of AGN; however, unlike optical emission, its origin is currently unclear. 
In contrast to X-ray binaries,  standard AGN accretion disk theory predicts disk temperatures too low to emit X-rays. 
The X-ray emission is thought by many to be generated by inverse Compton scattering of UV photons emitted from the disk by hot electrons in a corona above/around the disk (see review of Reynolds \& Nowak 2003).
However, both the geometry (e.g., its size,  shape, and position, e.g., height above the accretion disk) and the physics (e.g., the mechanism for heating the electrons) of the corona are under debate (Haardt \& Maraschi 1991; Haardt et al.\ 1994; Collin et al.\ 2003; Torricelli-Ciamponi et al.\ 2005; Trzesniewski et al.\ 2011).
Based on the Bayesian Monte Carlo microlensing analysis technique (Kochanek 2004;  Kochanek et al.\ 2007), recent observations in gravitational microlensing have constrained the X-ray emission size of quasars to be of order 10 $r_g$ (gravitational radius, $GM/c^2$, Blackburne et al.\ 2006;  Pooley et al.\ 2006; Morgan et al.\ 2008; Chartas et al.\ 2009; Dai et al.\ 2010; Blackburne et al.\ 2011), about one order of magnitude smaller than the optical emission region (Poindexter et al.\ 2008; Morgan et al.\ 2010; Mosquera et al.\ 2011). 
Chen et al.\ (2011, 2012a) detected energy-dependent X-ray microlensing in Q~2237+0305 and several other lenses for the first time, and their results suggest that the hard X-ray emission might come from regions smaller than the soft, while in RXJ~1131$-$1231 the effect is the opposite (Chartas et al.\ 2012).
Morgan et al.\ (2012)  found that the majority of the X-ray emission in QJ~0158$-$4325 comes from regions close to $6\, r_g,$ \ie in the vicinity of the last stable orbit of a Schwarzschild black hole.  

For distances less than a few gravitational radii, the flux and profile of both the X-ray continuum and metal emission lines (in particular, the \feka\ line) are strongly influenced by the gravitational field of the supermassive black hole at the center of an AGN. 
Due to the complexity of the Kerr metric (space-time of a black hole with angular momentum, or spin), much work has focused on either the analytic treatment of the geodesics in a Kerr metric (Kraniotis 2004, 2005; Hackmann et al.\ 2010) or on developing fast ray-tracing codes (Karas et al.\ 1992;  Dexter \& Agol 2009; Vincent et al.\ 2011). 
There has also been much work studying AGN X-ray spectra under strong gravity (Fabian et al.\ 1989;  Chen et al.\ 1989; Laor et al.\ 1989; Laor 1991;  Hameury et al.\ 1994; Bromley et al.\ 1997; Fabian \& Vaughan 2003; Beckwith \& Done 2004; Fuerst \& Wu 2004; Schnittman \& Krolik 2010).
The observed broad, skewed iron lines (e.g., MCG6$-$30$-$15, Tanaka et al.\ 1995) strongly suggest that the reflection component near the black hole contributes significantly to the total emissivity. 
Since the iron line profile strongly depends on the black hole spin, and the relativistic distortion of the iron emission line is much more easily measured than the broad X-ray continuum, most of these papers concentrated on the iron emission line. 
For example, Fuerst \& Wu (2004) studied the radiation transfer of emission lines in the Kerr space-time.
However, less work has been done on the continuum emission. 
It has been found that strong lensing of an accretion disk by the central supermassive black hole is strongly dependent on inclination angles (Cunningham 1975), and for very large inclination angles higher order images can contribute a significant fraction of the total luminosity (Beckwith \& Done 2005).
Polarization of the continuum emission near black holes was investigated in Connors et al.\ (1980) and is becoming an important tool in measuring the inclination angle of an accretion disk and the spin of its black hole (Dov$\rm\check{c}iak$ et al.\ 2004; McClintock et al.\ 2006; Li et al.\ 2009; Schnittman \& Krolik 2009; Dov$\rm\check{c}$iak et al.\ 2008, 2011).
Quasi--periodic oscillations (QPOs) have been observed in the X-ray light curves of black hole binaries and black hole candidates for more than a decade (Remillard \& McClintock 2006); however, no convincing QPO signal has been detected for AGN until very recently (Gierlinski et al. 2008; Middleton et al.\ 2009). 
High frequency QPOs  are often modeled in the context of strong gravity (the oscillation period is comparable to the Keplerian orbital period near the inner most stable orbit), and this can be used to accurately measure the black hole spin once the correct model of QPOs has been found (Wagoner et al.\ 2001; Schnittman 2005). 
The study of radiation transfer is crucial for these subjects, and Monte Carlo ray-tracing methods have been used extensively to solve (continuum or line) radiation transfer problems in AGNs (Agol \& Blaes 1996; Laurent \& Titarchuk 1999;  B$\rm\ddot{o}$ttcher et al.\ 2003; Dolence et al.\ 2009; Schnittman \& Krolik 2010). 
However, modeling of the AGN X-ray continuum emission was hampered by the lack of an accurate estimate of the size of the emission region.
We study the influence of the strong field of the massive spinning black hole centering the AGN on the X-ray continuum emission using the most recent constraints from quasar microlensing observations and our ray-tracing code. 
Despite the fact that there are nicely written public codes for ray-tracing in Kerr space-time available (e.g., Dexter \& Agol 2009), we develop our own ray-tracing code (specially designed for backward ray-tracing, see \textsection\ref{sec:BRT})  with a future goal to solve the radiation transfer equation in Kerr space-time using the characteristics method and operator perturbation algorithm (Hauschildt \& Baron 2006; Chen et al.\ 2007; Baron et al.\ 2009a,b).

In this paper, we study the inclination angle dependence of the flux profiles in the X-ray and the optical continuum of AGNs and their ratios caused by strong gravity, providing a reference for future usage. 
The X-ray-to-optical flux ratio, conventionally characterized by a parameter $\alpha_{\rm ox}$ (Tananbaum et al.\ 1979), is empirically measured in many AGNs.  
Our inclination dependent $\alpha_{\rm ox}$ with the Kerr lensing and projection effects taken into account will provide a baseline model for interpreting the observed $\alpha_{\rm ox}$ in AGNs.
For example, assuming that quasars are observed at randomly distributed solid angles, if there exists strong inclination angle dependence for the X-ray-to-optical flux ratios caused by the strong gravity field, the flux ratios will be observed with larger fluctuations than the intrinsic ones.
The mean value, scatter, and shape of the distributions might be changed by the strong gravity (see \textsection\ref{sec:observation}).  
Therefore, this can provide a test for different corona geometries, combined with a large number of measurements of \aox\ from different AGNs.

Based on the unification model (Antonucci 1993), AGN can be classified as Type I (normal) or Type II (obscured, possibly by a torus) with $\sim$50\% Type II AGN locally at low redshifts (e.g., Simpson 2005; Maiolino et al.\ 2007; Gilli et al.\ 2007).
There are suggestions that the fraction of Type II AGN decreases with luminosity or increases with redshift (Hasinger 2008; Treister et al.\ 2009; Bianchi et al.\ 2012).
Because emission of Type II AGN is strongly modified by Compton-thick absorption, which makes it difficult to study the intrinsic $\alpha_{\rm ox},$ we do not compare our model predictions to Type II AGN.
In Type I quasars, a significant fraction (20-40\%) are broad absorption line (BAL) quasars (e.g., Dai et al.\ 2008; Shankar et al.\ 2008).
These BALs are only moderately obscured, and several studies show that their intrinsic $\alpha_{\rm ox}$ are consistent with non-BALs (Green \& Mathur 1996;  Brandt et al.\ 2000; Green et al.\ 2001; Gallagher et al.\ 2002).
Therefore, we compare our model predictions with $\alpha_{\rm ox}$ measurements in Type I quasars including BALs.
If the X-ray-to-optical magnification ratio $R_{\rm X/O}$ is different between normal quasars and BALs, then there might be bias in the current estimate of  BAL absorption caused by ignoring the effect of strong gravity.
We investigate these ideas in this paper. 

In \textsection\ref{sec:Ray-Tracing} we introduce the ray-tracing code for Kerr space-time developed in this paper. 
In \textsection\ref{sec:basics} we present the basic formalism for studying strong lensing of Kerr space-time. 
In \textsection\ref{sec:optical} we study the influence of strong gravity on the optical emission.
In \textsection\ref{sec:Xrays}  we study strong lensing of the X-ray emission assuming three X-ray corona geometries.
In \textsection\ref{sec:observation} we compare the lensed $\alpha_{\rm ox}$ of normal/BAL quasars with observations, and estimate possible errors in current measurements of BAL absorption caused by ignoring effects of the Kerr strong gravity.
In \textsection\ref{sec:conclusion} we draw our conclusions.
Readers not interested in mathematical details are invited to skip \textsection\ref{sec:Ray-Tracing}, \textsection\ref{sec:BRT}, and all or part of \textsection\ref{sec:Xrays}.

%-------------------------------------------
\section{Ray-Tracing In Kerr Space-time}\label{sec:Ray-Tracing}

A Kerr black hole is fully characterized by two parameters, \ie mass M and angular momentum $a.$ 
In Boyer-Lindquist coordinates (Boyer \& Lindquist 1967), the Kerr metric (Kerr 1963) with signature $(-,+,+,+)$ can be written as
\bea
ds^2 %&=&-\frac{\rho^2\Delta}{\Sigma^2}dt^2 +\frac{\Sigma^2}{\rho^2}\sin^2\theta(d\phi-\frac{2aMr}{\Sigma^2}dt)^2+\frac{\rho^2}{\Delta}dr^2+\rho^2d\theta^2\cr
=-\alpha^2dt^2 +\frac{\Sigma^2}{\rho^2}\sin^2\theta(d\phi-\omega dt)^2+\frac{\rho^2}{\Delta}dr^2+\rho^2d\theta^2\eea where  $t$, $r$, $\theta,$ and $\phi$ are the four independent coordinate variables, and 
\bea\label{omega}
\rho^2&\equiv& r^2+a^2\cos^2\theta\cr
\Delta&\equiv&r^2-2Mr+a^2\cr
\Sigma^2&\equiv& \rho^2\left[(r^2+a^2)+\frac{2Mr}{\rho^2}a^2\sin^2\theta\right]\cr
\alpha^2&\equiv& \frac{\rho^2\Delta}{\Sigma^2}   \cr
\omega&\equiv& \frac{2aMr}{\Sigma^2}.
\eea 
The off-diagonal terms of the Kerr metric are characterized by $\omega$, which represents the dragging of the inertial frame.
A photon's geodesic motion can be described by a Hamiltonian ${\cal H}(x^a, p_a)$ defined on an eight-dimensional phase space (the index $a$ runs from 0 to 3), 
\bea\label{Hamilton}
{\cal H} &\equiv &\frac{1}{2}g^{ab}(x)p_ap_b \cr
&=& -\frac{1}{2\alpha^2}p_t^2-\frac{2Mar}{\rho^2\Delta}p_t p_\phi+ \frac{\Delta}{2\rho^2}p_r^2+\frac{1}{2\rho^2}p_\theta^2+\frac{\Delta-a^2\sin^2\theta}{2\rho^2\Delta\sin^2\theta} p_\phi^2,
\eea where $g^{ab}$ is the inverse metric tensor and $p_a$ is the covariant photon 4-momentum vector in the coordinate basis. 
To obtain photon trajectories in the Kerr space-time,  we solve the group of eight first order Hamilton equations 
\bea\label{H_equation}
\frac{dx^a}{d\xi}&=&\frac{\partial {\cal H}}{\partial p_a}  \cr% =g^{ab}p_b=p^a,\cr
\frac{dp_a}{d\xi}&=&-\frac{\partial {\cal H}}{\partial x^a}%=-\frac{1}{2}g^{bc}_{\>\>\>, a}p_b p_c.
,\eea where $\xi$ is the affine parameter  (refer to Chandrasekhar 1983 for details), instead of the four second order geodesic equations 
\be
\frac{d^2x^a}{d\xi^2}+\Gamma^{a}_{\>bc}\frac{dx^b}{d\xi}\frac{dx^c}{d\xi}=0,
\ee where $\Gamma^a_{\>bc}$ are the Christoffel symbols. 
The number of equations reduces to six since $p_t$ and $p_\phi$ are motion constants, and we give them in the Appendix. 
These six first order differential equations are much easier to solve numerically than the four second order differential equations. 
There is another motion constant discovered by Carter (1968),
\be
{\cal C}= p^2_\theta+\cos^2\theta\left(\frac{p^2_\phi}{\sin^2\theta}-a^2p_t^2\right),
\ee which we use as an independent check for our ray-tracing code (the maximum relative error of $\cal C$ allowed in the ray-tracing is less than $10^{-7}$). 
We solve Eqs.\,(\ref{H_equation}) numerically using a fifth-order Runge-Kutta algorithm with adaptive step size control (see e.g., Press et al.\ 1992). 

%The algorithm can be easily find in literature, e.g. Press et al.\ (1992), and is therefore not repeated here. 
%in which the most time expensive part  is the iteration scheme itself, not the cost in tracing the geodesics.     

Let $\nu_{\rm o}$ and $\nu_{\rm e}$ be the photon frequency measured by a distance observer and  an observer co-moving with the light source.
We  define the redshift factor
\be
g\equiv\frac{\nu_{\rm o}}{\nu_e} =\frac{(-u^ap_a)_{\rm o}}{(-u^ap_a)_{\rm e}},  % =\frac{1}{1+z},
\ee where  $u^a_{\rm o}$  and $u^a_{\rm e}$ are the four-velocity of the distant observer  and the source. 
For the distant observer, we choose 
\be
u^a=\frac{1}{\alpha}(1,0,0,\omega), \>\>\>  u_a=(-\alpha, 0, 0, 0)
\ee (the so-called zero-angular-momentum observer). 
An advantage of this choice is that the collection $\{u^a, \partial_r,\partial_\theta,\partial_\phi\}$ forms an orthogonal basis and therefore the photon's spatial direction can be easily specified with respect to  $\{\partial_r,\partial_\theta,\partial_\phi\}$, whereas  $\{\partial_t,\partial_r,\partial_\theta,\partial_\phi\}$ does not form an orthogonal basis (Bardeen et al.\ 1972). 
The four-velocity of the light emitter is model-dependent.  
We assume that the source is circling around the z-axis, \ie  
\be\label{u_K}
u^a=\gamma(1, 0, 0,\Omega). 
\ee 
The selection of $\Omega$ is not arbitrary, but constrained by the causality condition $u^au_a=-1.$ 
If the source is on the accretion disk (e.g., optical emission), we assume a prograde\footnote{Here prograde means that ${\bf a}\cdot{\bf J}>0$ where $\bf a$ is the (vector) spin of the black hole, and ${\bf J}$ is the angular momentum of the accretion disk.} Keplerian flow for the fluid (Bardeen et al.\ 1972), \ie
\be\label{Omega_K}
\Omega=\Omega_K\equiv\frac{M^{1/2}}{r^{3/2}+a M^{1/2}}.
\ee 
If the source is not in the equatorial plane (e.g., the X-ray ball model in \textsection\ref{sec:Xball}), we use the profile of Ruszkowski \& Fabian (2000), 
\be\label{RF_Omega}
\Omega =\left(\frac{\theta}{\pi/2}\right)\Omega_K+\left[1-\left(\frac{\theta}{\pi/2}\right)\right]\omega,
\ee where $\theta$ is the spherical polar angle in the Boyer-Lindquist coordinates (an interpolation between $\Omega_K$ and $\omega$ of the frame dragging, see Eq.~(\ref{omega})).

%---------------------------------------
\section{Basics of Strong Lensing in Kerr Space-Time}\label{sec:strong-lensing}\label{sec:basics}

Strong lensing effects of a Kerr black hole on the accretion disk and X-ray corona differ from standard linear gravitational lensing theory in a few important aspects.  
First, linear lensing theory requires a very small bending angle, \ie $GM/r_0\ll 1$  where $M$ and $r_0$ are the lens mass and the impact vector, respectively. 
This assumption breaks down for accretion disks whose inner cutoff can be as close as a few gravitational radii from the lensing black hole.  
Second, the source, lens, and observer are far away from each other in the linear lensing theory, whereas for Kerr lensing, there is no well-separated lens and source planes, \ie the source is very close to the lens.
Third, the space-time in the source plane is assumed to be flat in the linear theory, whereas for Kerr lensing, this is not true. 
Last, since both the source and the observer are far away from the lens in the linear theory, there is no gravitational redshift caused by the lens, and the redshift  comes solely from cosmology. 
This is incorrect for Kerr lensing. 
Since photons are emitted by sources moving with relativistic speeds in strongly curved space-time, both the Doppler and gravitational redshifts have to be taken into account in Kerr lensing.
Consequently, there is no simple lens equation relating the source and image positions through a single bending angle (the bending happens continuously, instead of at a single point in the lens plane as assumed in the linear lensing theory). 
We identify the source position for each ray arriving at the observer through backward ray-tracing in the Kerr space-time (see \textsection\ref{sec:BRT}).  
The redshift effect is very important  for computing observable quantities from the accretion disk around the supermassive black hole.   

By definition, the monochromatic flux $F_{\nu_{\rm o}}$ measured by an observer O sitting in a (asymptotically) flat space-time is   
\be
F_{\nu_{\rm o}}\equiv\oint{I_{\nu_{\rm o}}\mu d\Omega_{\rm o}}= \int{I_{\nu_{\rm o}} d\Omega_{\rm o}},
\ee where $d\Omega_{\rm o}$ is the differential solid angle measured at the observer, $\mu$ is the angle cosine with respect to the normal of the detector window (see Figure~\ref{fig:raytracing}), and can be safely dropped for sources very far away from the observer, \ie the source is observed with a tiny solid angle. 
In standard lensing theory, there are no frequency redshifts between the source and observer (besides the cosmological redshift), therefore $I_\nu$ is a conserved quantity along the light path in case of no scattering and absorption, thus
\be
F_{\nu_{\rm o}}\equiv \int{I^{\rm obs}_{\nu_{\rm o}}(\mbox{\boldmath$\theta$})d\Omega_{\rm o}}= \int{I^{\rm source}_{\nu_{\rm o}}(\mbox{\boldmath$\beta$}) d\Omega_{\rm o}},
\ee where $\mbox{\boldmath$\theta$}$ and $\mbox{\boldmath$\beta$}$ are the image and source positions, respectively. 
Since the space-time in both the source and lens plane are assumed to be flat, the integral over solid angle can be easily transformed into 2-D surface integrals in the lens or source plane. 
To compute the flux magnification, we simply compute the ratio of the image area to the source area in the lens plane,  \ie  
\be
{\cal A}^{-1}\equiv \det  \frac{\partial \mbox{\boldmath$\beta$} }{\partial\mbox{\boldmath$\theta$}}.
\ee  
This is equivalent to computing the ratios of the solid angles subtended by the image and the source at the observer.  
On the other hand, for Kerr lensing, we have 
\be\label{F_nu_defi}
F_{\nu_{\rm o}}\equiv \int{I^{\rm obs}_{\nu_{\rm o}}(\nhat)d\Omega_{\rm o}} = \int{g^3(\nhat) I^{\rm source}_{\nu_{\rm e}}(x^a(\nhat),p_a(\nhat)) d\Omega_{\rm o}},
\ee where $\nhat$ is the photon's 3-D direction at the observer, $(x^a,p_a)$ is the photon's position and momentum at the emitter (found through backward ray-tracing), $\nu_{\rm e}$ is the source frequency, and $g$ is the redshift factor defined earlier.
Clearly, the effect of Kerr lensing strongly depends on the redshift factor $g,$ and consequently on the four-velocity of the source. 
Solid angles (at the observer) and areas (in the lens or source plane) are often used interchangeably in linear lensing theory; however, in Kerr lensing theory, solid angles at the observer are more appropriate quantities to use since the space-time around the source and lens is strongly curved.\footnote{The familiar angular diameter distance defined in the FLRW cosmology, \eg $D_s$ of the source,  can not be used to compute the source area directly since it does not respect the space-time curvature at the source caused by the supermassive black hole.}

%------------------------------------------------------------------------------
\subsection{Models of Optical and X-Ray Emission}

We assume that the optical/UV emission comes from the accretion disk (Shakura \& Sunyaev 1973), and is Planckian, 
\be\label{Planck}
B(\nu,r)=\frac{2h}{c^2}\frac{\nu^3}{\exp\left[\frac{h\nu}{k T(r)}\right]-1}.   %=\frac{2h}{c^2}\frac{\nu^3}{\exp\left(\frac{r}{r_s}\right)^{3/4}-1}
\ee 
The effective temperature can be approximated as $T(r)\propto r^{-3/4}\kappa^{1/4}$ where $\kappa=1-\sqrt{r_{\rm ISCO}/r}$ ($r_{\rm ISCO}$ is the innermost stable circular orbit,  Bardeen et al.\ 1972).  
For convenience we set $\kappa$ to unity since this factor only influences the innermost region of the accretion disk which does not contribute significantly to the total optical flux.
Likewise we have ignored the general relativistic corrections to $T(r)$ in the innermost regions  (Novikov \& Thorne 1973).
%where $r_s$ is the  scale radius for frequency $\nu$  defined such that $h\nu=k T(r_s)$ where $k$ is the Boltzmann constant and $T$ is the temperature of the disk, and we have used $T\propto r^{-3/4}$ (Shakura \& Sunyaev 1973). 
From Eq.~(\ref{F_nu_defi}), the observed monochromatic flux $F_{\nu_{\rm o}}$ at frequency $\nu_{\rm o}$ is  
\bea\label{optical}
F_{\nu_{\rm o}}  &=& \int{g^3 B(\nu_{\rm e},r) d\Omega_{\rm o}}    %  \int{g^3 \frac{2h}{c^2}\frac{\nu_e^3}{\exp\left[\frac{h\nu_e}{kT(r)}\right]-1}d\Omega_{\rm o}}
=\int{ \frac{2h}{c^2}\frac{\nu_{\rm o}^3}{\exp\left[\frac{h\nu_{\rm o}}{gkT(r)}\right]-1}d\Omega_{\rm o}} 
=\int{ \frac{2h}{c^2}\frac{\nu_{\rm o}^3}{\exp\left[\frac{1}{g}\frac{T_o}{T(r)}\right]-1}d\Omega_{\rm o}}\cr
&=&\int{ \frac{2h}{c^2}\frac{\nu_{\rm o}^3}{\exp\left[\frac{1}{g}\left(\frac{r}{r_s}\right)^{\frac{3}{4}}\right]-1}d\Omega_{\rm o}}
=\int{B(\nu_{\rm o},\frac{r}{g^{4/3}})d\Omega_{\rm o}}.
\eea 
In the above derivations we have defined the so-called scale radius $r_s$ (not the Schwarzschild radius) by $h\nu_{\rm o} \equiv  k T_{\rm o} \equiv kT(r_s).$  
We assume that the accretion disk moves with the Keplerian flow, see Eq. (\ref{Omega_K}).  
We choose the inner cutoff of the optical emission disk at the innermost stable circular orbit, $r_{\rm ISCO},$ 
where $r_{\rm ISCO}=6\,r_g, \,4.23\,r_g$, and $1.24\,r_g$ for $a=0, \,0.5M,$ and $0.998M$ black holes, respectively (Bardeen et al.\ 1972). 
This last value corresponds to an extreme Kerr black hole (Thorne 1974). 

The spectrum of the X-ray emission of AGN is observed to follow a power-law. 
We therefore assume a simple power-law model for the  X-ray emission, 
\be\label{X-ray_profile}
I_\nu(\nu, r) \propto \frac{1}{r^n}\frac{1}{\nu^{\Gamma-1}},
\ee where $r$ is the radial coordinate of the source, $n$ specifies the steepness of the radial profile, and $\Gamma$ is the photon index. 
Using Eq.\,(\ref{X-ray_profile}), we also assume that the X-ray emission is isotropic in the local rest frame of the X-ray source.\footnote{The isotropy assumption is made for simplicity. However, the formalism developed in this paper does not require isotropic intensity profiles.} 
The observed monochromatic flux is 
\bea\label{F_nu}
F_{\nu_{\rm o}} %=\int{I_{\nu_{\rm o}}\mu d\Omega_{\rm o}}    %= \int{I_{\nu_{\rm o}} d\Omega_{\rm o}} 
=\int{g^3 I_{\nu_e}d\Omega_{\rm o}}=\int{\frac{g^3}{r^n\nu_e^{\Gamma-1}}d\Omega_{\rm o}}
=\int{\frac{g^{\Gamma+2}}{r^n\nu_{\rm o}^{\Gamma-1}}d\Omega_{\rm o}}
=\frac{1}{\nu_{\rm o}^{\Gamma-1}}\int{\frac{g^{\Gamma+2}}{r^n}d\Omega_{\rm o}},
\eea which follows a power-law with the same photon index $\Gamma$ and we have dropped the unimportant constant in Eq.~(\ref{X-ray_profile}). 
The strong lensing correction is contained in the last integral in the equation above, and the redshift factor $g$ depends only on the photon trajectory, or equivalently on the photon's arriving direction $\hat{\bf n}$ at the observer,  but not on its frequency because gravitational lensing is achromatic. 
Different radial profiles and sizes of X-ray emission regions give different strong lensing corrections.     
The observed bolometric flux (e.g.,  1--50 keV, in the observer's frame) is 
\bea
F\equiv\int_{\nu_{\rm min}}^{\nu_{\rm max}}{F_{\nu_{\rm o}}d\nu_{\rm o}}
=\int_{\nu_{\rm min}}^{\nu_{\rm max}}{\frac{1}{\nu_{\rm o}^{\Gamma-1}}d\nu_{\rm o}}\times\int{\frac{g^{\Gamma+2}}{r^n}d\Omega_{\rm o}}.
\eea  
Since the first integral in the above equation is independent of the second one, the strong lensing corrections to the integrated flux and the monochromatic flux are the same for the simple X-ray emission profiles considered in this paper. 

\begin{figure}
	\epsscale{0.8}
	\plotone{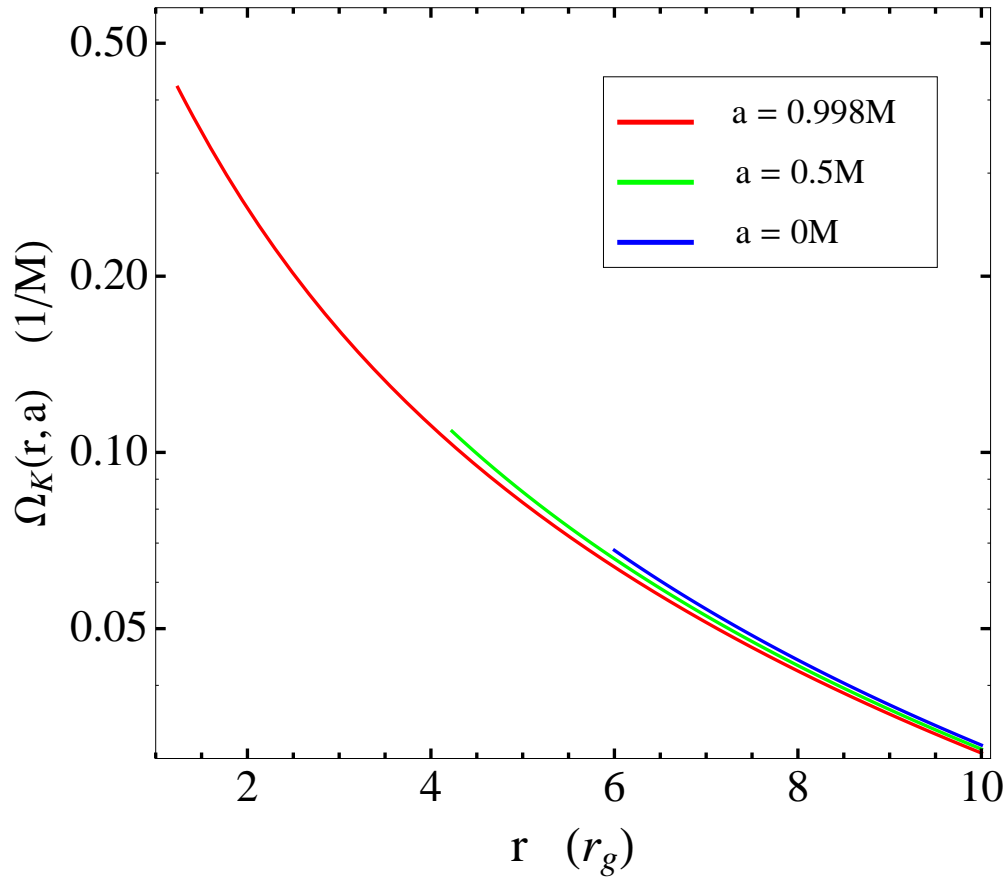}
	\caption{ Angular velocity $\Omega$ for the Keplerian flow (see Eq.~\ref{Omega_K}) as a function of the Kerr radial coordinate $r$ for the low X-ray disk model with radius $r_{\rm disk}=10\, r_g$ and inner cutoff at $r_{\rm ISCO }=6\,r_g,$ $4.23\,r_g,$ and $1.24\,r_g$ for  spin $a=$ 0, 0.5\,M, and 0.998 M, respectively.
		 \label{fig:Xray_Omega}} 
\end{figure}

We consider three different X-ray geometries:  a thin X-ray disk immediately above the accretion disk moving with Keplerian flow, see Figure~\ref{fig:Xray_Omega} (low X-ray disk model), a spherical corona around the black hole (X-ray ball model), and a thin disk above the accretion disk with different heights (high X-ray disk model).
For the low X-ray disk model, we choose the inner cutoff at $r_{\rm ISCO}.$ 
We choose no inner cutoff for the high X-ray disk model. 
For the X-ray ball and high X-ray disk model, we assume that the X-ray source is moving with the Ruszkowski \& Fabian (2000) profile, Eq.~(\ref{RF_Omega}), see Figures~\ref{fig:Xball_Omega} and \ref{fig:Xover_Omega}.
We take the photon index $\Gamma=2,$ which is normal for quasar X-ray spectra.
For each X-ray geometry, we evaluate three spin parameters: a = 0, 0.5\,M, and 0.998\,M. 

%---------------------------------------------------
\subsection{Backward Ray-Tracing}\label{sec:BRT}

%For flat space-time, $g\equiv 1$ and light paths are straight lines.
For flat space-time, light paths are straight lines, and there is no gravitational redshift.
Ignoring the special relativistic Doppler shifts and aberration effect, we have $g=1.$ 
Therefore, Eqs~(\ref{optical}) and (\ref{F_nu}) can be easily carried out as a surface integral over the accretion disk or the X-ray source, \ie
\be\label{F_O_Euc}
F_{\nu_{\rm o}}^{\rm O (flat)}=\frac{2\pi\cos\theta}{r_{\rm obs}^2}\int{B(\nu_{\rm o}, r)rdr},
\ee and
\bea\label{F_nu_Euc}
F_{\nu_{\rm o}}^{\rm X (flat)}=\frac{1}{\nu_{\rm o}^{\Gamma-1}}\int{\frac{d\Omega_{\rm o}}{r^n}}
%= \frac{1}{\nu_{\rm o}^{\Gamma-1}}\int{\frac{1}{r^n}\frac{d {\cal A}_e^{\perp}}{r_{\rm obs}^2}}
= \frac{1}{\nu_{\rm o}^{\Gamma-1}}\frac{1}{r_{\rm obs}^2}\int{\frac{d{\cal A}_e^{\perp}}{r^n}},
\eea where $\theta$ is the inclination angle of the disk, $r_{\rm obs}$ is the radial coordinate of the observer, and $d{\cal A}_e^{\perp}$ is the differential (X-ray) source area orthogonal to the line of sight.
For curved space-time with relativistic flows, Eqs.~(\ref{optical}) and (\ref{F_nu}) need to be computed numerically. 

\begin{figure}
	\epsscale{0.9}
	\plotone{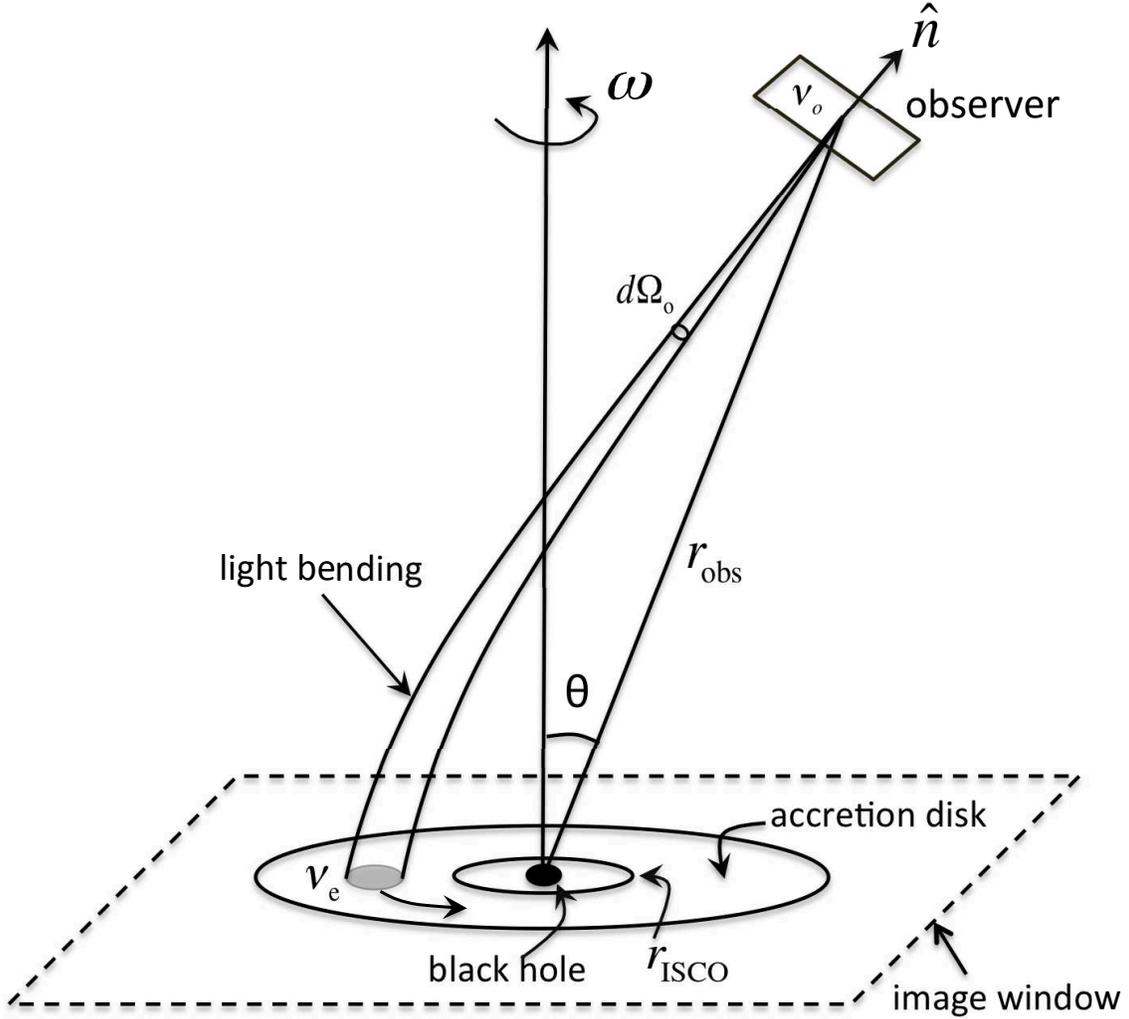}
	\caption{Schematic plot for backward raytracing. 
	 $\theta$ is the inclination angle.
	 $r_{\rm ISCO}$ is the innermost stable circular orbit.
	 $\mbox{\boldmath$\omega$}$ is the dragging of the inertial frame.
	 The dashed parallelogram is the intersection of the pencil of light beam (shooting backward from the observer) with the equatorial plane whose projection along the line of sight is the image window.
	 The observer is sufficiently far away from the black hole, such that the space-time is nearly flat, and is close enough such that the observer has the same cosmological redshift. 
	 The accretion disk size is greatly exaggerated with respect to the observer distance $r_{\rm obs}.$
	  A photon is emitted with frequency $\nu_{\rm e}$  (measured in the local rest frame) from the disk and observed with gravity and Doppler red/blue shifted frequency $\nu_{\rm o}$ along direction $\nhat$ measured by the so-called  zero-angular-momentum-observer.  
	The light path is bent by the strong gravity produced by the central supermassive black hole.     
            \label{fig:raytracing}} 
\end{figure}

We choose a pencil beam with a rectangular cross-section, large enough to contain all rays arriving at the observer (Figure~\ref{fig:raytracing}), around the central light ray toward the black hole, and divide the solid angle space into a uniform grid.  
Each pixel, labeled by $(i,j)$, corresponds to an image area $dA_{i,j}=d\Omega_{i,j}r^2_{\rm obs}.$ 
Since we have used a uniform rectangular grid, $d\Omega_{ij}=d\Omega$ is constant for each pixel. 
Therefore, to compute the image area, we need only count the number of pixels whose light rays end at the accretion disk. 
To compute the observed flux, we weight pixels  by the integrands in  Eqs~(\ref{optical}) and (\ref{F_nu}) for optical and X-ray emission, respectively.  
Eqs~(\ref{optical}) and (\ref{F_nu}) are written as integrals over solid angle at the observer; however, we emphasize that the integrands are evaluated at the emitter (near the black hole), and each piece of solid angle $d\Omega_{\rm o}(\nhat)$ is mapped to a photon's initial phase space coordinate $(x^a,p_a)$ on the accretion disk through backward ray-tracing. 
For example, if a photon from the optical accretion disk arrives at the observer along direction $\hat{\bf n},$ we find its $r$ coordinate on the accretion disk and redshift factor $g$ through backward ray-tracing (Figure~\ref{fig:raytracing}). 
If $g<1$ (redshifted), then we sample the temperature of the blackbody radiation at radius $r/g^{4/3}$ ($> r$,  $T(r/g^{4/3})<T(r)$).  
 This approach takes into account gravitational light deflection, frequency redshift, and area distortion automatically. 
We do not use the popular transfer function method (Cunningham 1975), which only considers light emitted on the accretion disk (the equatorial plane), since we allow X-rays to be emitted outside of the equatorial plane. 

%(e.g., for an accretion disk, $d{\cal A}^{\perp}_e =d{\cal A}_e\cos\theta$ where $d{\cal A}_e$ is the area element on the disk and $\theta$ is the inclination angle of the accretion disk). 
% This turns the solid angle integral at the observer into a surface integral over the source (e.g., the accretion disk). 

%To compute the strong gravitationally lensing corrected specific  flux,  for each pixel hitting on the accretion disk, I record its $r$ coordinate and the redshift factor $g$ and using Eq.\,(\ref{F_nu}). For the two cases with no light bending (but with flat or curved disk), I use Eq.\,(\ref{F_nu_Euc}) with $d{\cal A}= 2\pi rdr$ or using Eq.\,(\ref{dA_kerr}). 

%--------------------------------------------------------------------------------------------------------------------------
\subsection{Strong Lensing Correction to the X-Ray To Optical Flux Ratio}\label{sec:aox}

Let $M_{\rm O}$ and $M_{\rm X}$ be the strong lensing magnifications of the optical and X-ray emission, then the ratio between the observed monochromatic flux in the optical and X-ray band can be written 
%\be\label{modulation}
%R_{\rm X/O} \equiv \frac{F^{\rm X}_{\nu_{\rm o}}}{F^{\rm O}_{\nu'_{\rm o}}} = \frac{M_{\rm X}}{M_{\rm O}} \frac{F^{\rm X(flat)}_{\nu_{\rm o}}}{F^{\rm O (flat)}_{\nu'_{\rm o}}}
%\equiv R_{\rm X/O, lens}R_{\rm X/O, proj}  
%\ee
\be\label{modulation}
\frac{F^{\rm X}_{\nu_{\rm o}}}{F^{\rm O}_{\nu'_{\rm o}}}
= \frac{M_{\rm X}}{M_{\rm O}} \frac{F^{\rm X(flat)}_{\nu_{\rm o}}}{F^{\rm O (flat)}_{\nu'_{\rm o}}}
\equiv R_{\rm X/O}^{\rm (lens)}R_{\rm X/O}^{\rm (proj)}  \left.\frac{F^{\rm X}_{\nu_{\rm o}}}{F^{\rm O }_{\nu'_{\rm o}}}\right|_{\theta=0}^{\rm (flat)}
\equiv R_{\rm X/O}\left.\frac{F^{\rm X}_{\nu_{\rm o}}}{F^{\rm O }_{\nu'_{\rm o}}}\right|_{\theta=0}^{\rm (flat)} 
\ee  where we have defined the lensing magnification ratio,
\be
R_{\rm X/O}^{\rm (lens)}\equiv \frac{M_{\rm X}}{M_{\rm O}}, 
\ee the (Euclidean) projection area ratio $R_{\rm X/O}^{\rm (proj)}$ (with respect to the face on case), and the total ratio
\be\label{RXO}
R_{\rm X/O}\equiv R_{\rm X/O}^{\rm (lens)}R_{\rm X/O}^{\rm (proj)}  
\ee  which contains all inclination dependence (lensing plus projection). 
%\be
%R_{\rm X/O, proj} \equiv \frac{F^{\rm X(flat)}_{\nu_{\rm o}}}{F^{\rm O (flat)}_{\nu'_{\rm o}}}.
%\ee
If  the X-ray and optical emission regions are different in geometrical size, surface brightness profile, or are located at different places with respect to the central black hole, strong gravitational lensing of the Kerr black hole should magnify or demagnify these two emission sources differently.  
In particular, the inclination angle dependence of the magnification should be different for X-ray and optical emission, and the magnification ratio $R_{\rm X/O}^{\rm (lens)}$ should be a function of $\theta$. 
Among the three X-ray geometries considered, $R_{\rm X/O}^{\rm (proj)} = 1$ for the low and high X-ray disk model because the factor $\cos\theta$ is canceled when considering both the optical and X-ray emission, and $R_{\rm X/O}$ purely reflects the differential lensing effect.
For the X-ray ball model, $R_{\rm X/O}^{\rm (proj)} = (1+\cos\theta)/2\cos\theta$ is inclination-dependent (see \textsection\ref{sec:Xball}), where the projected X-ray emission area is non-vanishing at large inclination angles.
Thus, $R_{\rm X/O}$ is affected by both lensing and projection effects for the X-ray ball model.

We investigate strong lensing of optical and X-ray emission by Kerr black holes in the next two sections, and pay particular attention to the lensing magnification ratios $R_{\rm X/O}^{\rm (lens)}(\theta).$
In \textsection\ref{sec:observation} we compare the strong lensed modulated $\alpha_{\rm ox}$ with observations.

%------------------------------------------------------------------------------------------
\section{Strong Lensing of the Optical Emission}\label{sec:optical}

Because of uncertainties in parameters such as the black hole mass and the accretion efficiency, the scale radius $r_s$, or the half light radius (\ie radius of the circular region centering the peak brightness of the emission and containing half the total flux at a given wavelength) $r_{\rm half}=2.45\, r_{\rm s}$, cannot be precisely determined from the accretion disk theory.
Here, we use the empirical results from recent quasar microlensing observations (Kochanek 2004; Pooley et al.\ 2006; Morgan \etal 2008; Eigenbrod et al.\ 2008; Chartas \etal 2009, Dai \etal 2010, Blackburne \etal 2011; Blackburne et al.\ 2012), and assume that $r_{\rm half}=100 \, r_g,$ and take $r_s =41\, r_g.$ 
We choose the outer boundary of the optical disk at $r_{\rm disk} = 200\, r_g$  which is 5 times the scale radius $r_s,$ and contains $\sim$$80\%$ of the monochromatic flux. We ignore the lensing correction for regions with $r> 200\, r_g,$ and assume that the flux magnification ratio is 1 beyond $r_{\rm disk}.$ 
Consequently, if the flux magnification is $\rm M_{<200\,r_g}$ within $r_{\rm disk},$ the total flux magnification $\rm M_{O}$ is then $\rm M_{O}=0.2+0.8\,M_{<200\,r_g}.$       

We plot the strong lensing distortion to the image area and specific flux $F_{\nu_{o}}$ in Figure~\ref{fig:optical_mag} for three different spin parameters, a = 0, $0.5 M$, and $0.998 M$.  
Since the size of the optical emission region is relatively large (of an order $100\,r_g$), and the intensity profile (blackbody radiation with $T\propto r^{-3/4}$, see Eq.~(\ref{Planck})), is not very steep, the distortion in either the image area or flux is not significant (close to 1) for AGNs observed nearly face on, and the area/flux magnification curves $M_{\rm O}(\theta)$ are flat until the inclination angle $\theta$ becomes large ($\gtrsim75^\circ$, i.e., observed near edge on).  
Since the black hole spin $a$ contributes only in higher orders (e.g., $\omega\sim aM/r^3$), we do not see a significant difference in the flux or image area distortion for different spin parameters for optical emission. 
If we increase the disk size $r_{\rm disk}$, the magnification of image area $M^{\rm area}_{\rm O}$ at all inclination angles will eventually approach one, but the magnification in flux $M^{\rm flux}_{\rm O}$ changes only slightly, since the majority of the emission is enclosed within $r_{\rm disk}$ and the effect of gravity is weak outside of $r_{\rm disk}.$  

%For optical emission we considered only the primary image of the disk and did not allow photons to pass through the equatorial plane, \ie we assume that the accretion disk is opaque. 
%Allowing higher order images is numerically straightforward, but physically unrealistic. 
%To test our ray-tracing code, we allowed secondary images when doing the ray-tracing for the X-ray disk (see next section) and found that the contribution of the secondary images to observed flux, or image area is insignificant except for very large inclination angle $\theta$ or very small sources (see Figure \ref{fig:secondary_image}).   

%------------------------------------------------------------------
\section{Strong Lensing of the X-Ray Emission}\label{sec:Xrays}

In this section we consider three different X-ray models:  a thin X-ray disk immediately above the accretion disk (the low X-ray disk model), a spherical corona around the black hole (the X-ray ball model), and a thin disk above the accretion disk with different heights (the high X-ray disk model).
The results for these X-ray models as well as the optical emission model discussed in \textsection\ref{sec:optical} are summarized in Table~\ref{tab:Flux_Mag}.

%-------------------------------------------------------------------------------
\subsection{Low X-Ray Disk Model}\label{sec:Xray}

\begin{figure}
	\epsscale{1.0}
	\plotone{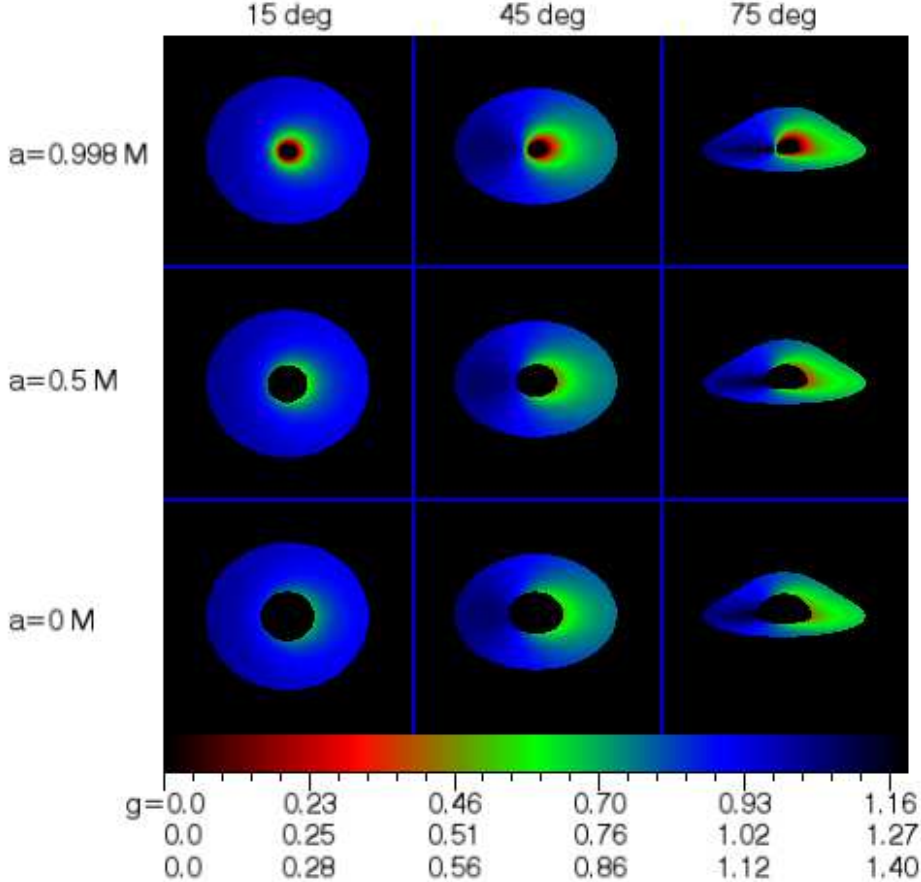}
	\caption{ Lensed and redshifted images for a (low) X-ray disk of radius $r_{\rm disk}= 20\,r_g$ moving with the Keplerian flow.   
	The top, middle, and bottom rows show the results for spins $a=0.998\,M,$ 0.5\,M, and 0 (corresponding to inner cutoff at $r_{\rm ISCO}=1.24\,r_g,$ $4.23\,r_g,$ and $6\,r_g$). 
	The first, second, and last columns are for inclination angles $\theta=15^\circ,$ $45^\circ,$ and $75^\circ,$ respectively. 
	The observer is at $r_{\rm obs} = 10^6\, r_g.$  
	The colorbar shows the redshift factor $g\equiv \nu_{\rm o}/\nu_{\rm e}$ for inclination angles $\theta=15^\circ$ (the first row), $45^\circ$ (the second row) and $75^\circ$ (the last row), respectively. 
	A disk observed at a large inclination angle spans a larger redshift interval.  
	The source on the left hand side is moving toward the observer (Doppler blueshifted), and the source on the right hand side is receding from the observer (Doppler redshifted). 
	The image of an optical emitting disk is similar (but larger) since they are both in the equatorial plane and move with the Keplerian flow.    
	 \label{fig:Xray_image}} 
\end{figure}

We first consider a simple model where X-rays are emitted by a tenuous disk immediately above the quasar accretion disk and this X-ray disk moves with the same Keplerian flow as the optical emission disk.
This is partially motivated by the so-called ``sandwich'' corona model (Haardt \& Maraschi 1991, 1993; Ruszkowski \& Fabian 2000).  
We choose the outer and inner cutoff at $r_{\rm disk}=10\, r_g$ and $r_{\rm in}= r_{\rm ISCO},$ respectively. 
%Similar to the optical emission case, we test  zero, moderate, and extreme spin case  ($a=$ 0, 0.5 M, and 0.998 M).
For the intensity profile,  we use  Eq.\,(\ref{X-ray_profile}), with $n= 3$ (\ie $I_{\nu}\propto r^{-3},$ following the optical profile) or $n=0$ (no radial dependence). 
For the case of an extreme Kerr black hole ($a=0.998\,M$), we also tried the case where $r_{\rm disk}= 5\, r_g,$ which is not realistic for $a=0$ or 0.5 M, considering the inner cutoff $r_{\rm ISCO}$ we choose.
We show the lensing distorted and redshifted images of such an X-ray disk in Figure~\ref{fig:Xray_image} for observers at $\theta =15^{\circ},$ $45^\circ,$ and $75^\circ$ (from nearly face on to nearly edge on), and three spins, $a = 0, 0.5$, and 0.998\,M. 
All images in Figure~\ref{fig:Xray_image} are distorted as results of the lensing effect.  Disks viewed at large inclination angles show stronger image distortion, and black holes with larger spins produce larger redshifts or blue-shifts.
Since the optical accretion disk (\textsection\ref{sec:optical}) is moving with the same Keplerian flow, the images for an optical disk are very similar, except that the image sizes are larger.  

\begin{figure}
	\epsscale{0.7}
	\plotone{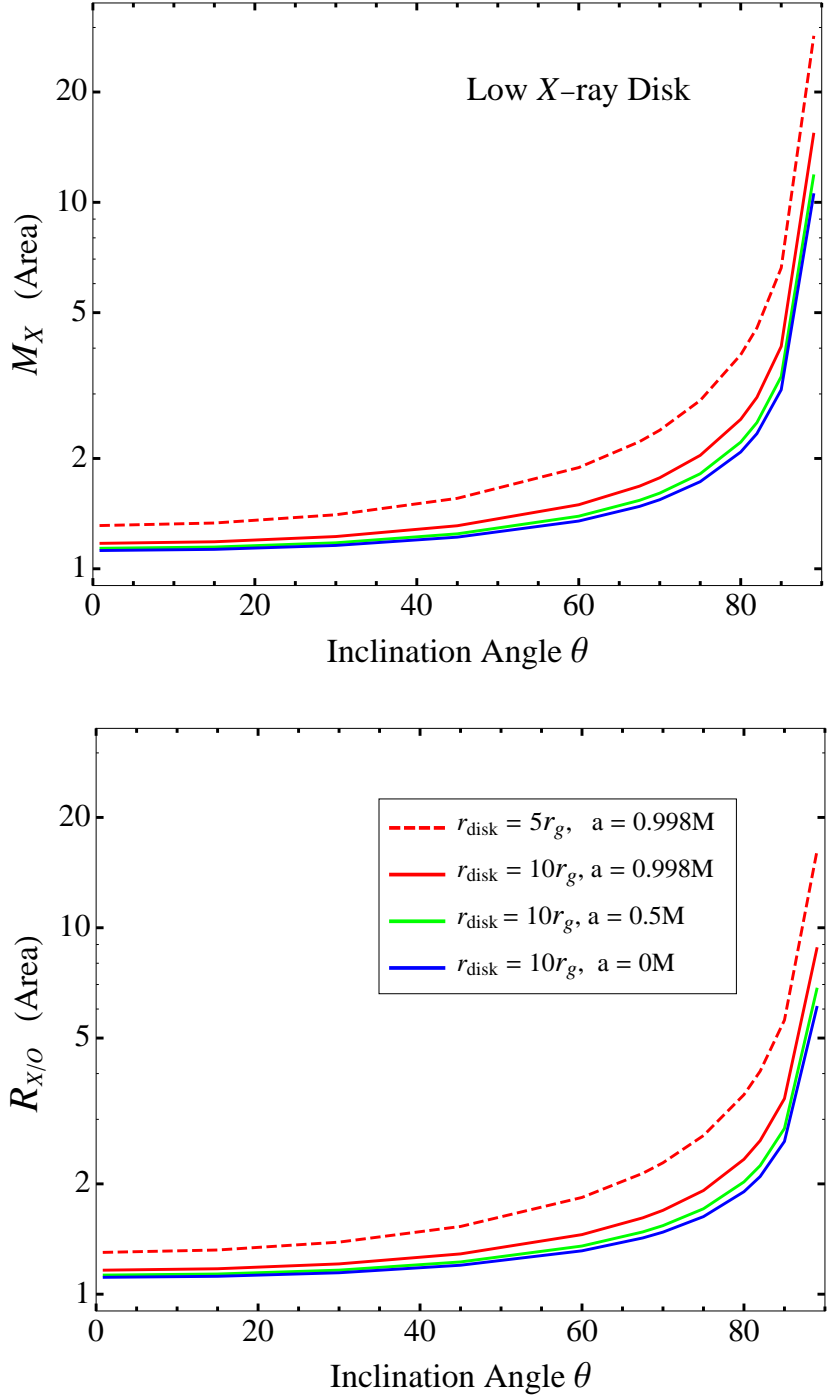}
	\caption{Inclination angle dependence of the strong lensing magnification of the image area of an X-ray disk.  
	We assume that X-rays are emitted from a thin disk immediately above the disk with $r_{\rm disk}=10\, r_g$ and $r_{\rm inner}= r_{\rm ISCO}$ for a = 0, 0.5\,M, and 0.998\,M. 
	For the $a =0.998 M$ case, we also considered $r_{\rm disk}= 5\,r_g$ (red dashed line). 
            \label{fig:Xray_Area_ratio}} 
\end{figure}

In Figure~\ref{fig:Xray_Area_ratio} we show the amplification of the X-ray image area and its ratio with respect to the optical image area for each case. 
%In the top panel we show the numerical values of the amplification in image area, $M_{\rm X},$ and in the bottom we compute the ratio between the amplification of the X-ray disk area and that of the optical disk, \ie $R_{\rm X/O}\equiv M_{\rm X}/M_{\rm O},$ and plot it as a function of inclination angle $\theta.$ 
The curves in the two panels resemble each other at $\theta\lesssim75^\circ$ since the optical area magnifications, $M_{\rm O}(\theta)$, are nearly flat and very close to $1$ in this range of inclination angles (Figure \ref{fig:optical_mag}).  
At larger inclination angles, the optical disk area is also significantly magnified (Figure \ref{fig:optical_mag}), and thus the corresponding curves in the two panels deviate from each other.  
The X-ray area amplification increases with inclination angles because more photons pass near the black hole at larger inclination angles, which are more strongly affected by lensing. 
The area amplification ratios also increases with inclination angles because the X-ray emission region is much smaller, resulting in stronger lensing. 

%In Figure~\ref{fig:Xray_flux} we show the inclination angle dependence of the observed flux (up to a scaling constant, which is the same for all cases) in the strong field of Kerr space-time. 
%In the presence of a strong gravitational field, we obtain a non-monotonic $F_\nu(\theta)$ curves for our  X-ray disk model.  
%In particular, for an extreme black hole with a steep radial profile ($n=3$), the observed $F_\nu$ increases with inclination angle $\theta$ for both choices of $r_{\rm disk}$. 

\begin{figure}
	\epsscale{0.9}
	\plotone{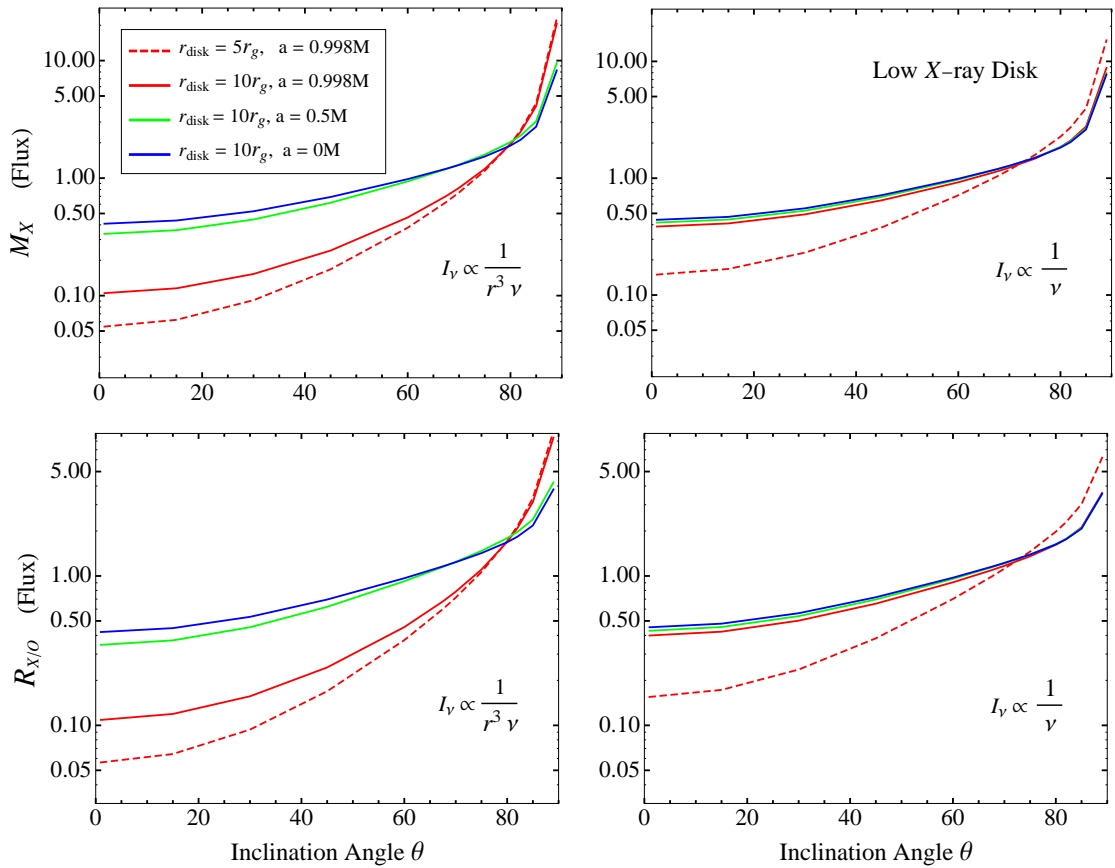}
	\caption{The inclination angle dependence of the strong lensing magnification of the X-ray flux for the low X-ray disk model. 
%	We assume that X-rays are emitted from a thin disk immediately above the optical emission (accretion) disk.  
	The first and second rows plot the magnification $M_{\rm X}$ and the X-ray-to-optical magnification ratio $R_{\rm X/O},$ respectively. 
	The projection effect (the $\cos\theta$ factor) cancels out for the X-ray disk model, and consequently the inclination-dependence of $R_{\rm X/O}$ is purely general relativistic.
%	$R_{\rm X/O}$ can change by a factor of $\sim$10 from low to moderate inclination angles, and another factor of $\sim$10 for high inclination angles as shown in the bottom left panel. 
	$R_{\rm X/O}$ depends on the inclination angle $\theta,$ and the dependence is stronger for smaller X-ray disks, larger spins, or steeper radial profiles.
	For the case $a=0.998\,M,$ $r_{\rm disk}=10\,r_g$, and $I_\nu\propto r^{-3},$  $R_{\rm X/O}$ increases by a factor of 10 when $\theta$ increases from $15^\circ$ to $75^\circ.$
	For the nearly edge on case, $R_{\rm X/O}$ can increase by another factor of $10$ in extreme cases with small X-ray emission sizes and steep profiles (see the bottom left panel). 
	 \label{fig:Xray_flux_ratio}} 
\end{figure}

In Figure~\ref{fig:Xray_flux_ratio} we show the magnification of the observed flux $F_\nu$ as a function of inclination angle. 
For small inclination angles, the Doppler effect is unimportant, and the gravitational redshift reduces the observed specific flux through the invariance of $I_\nu/\nu^3$.  
For large inclination angles,  the amplification of image area increases rapidly (with $\theta$) because of strong lensing (Figure~\ref{fig:Xray_Area_ratio}), and Doppler blue-shifted regions (approaching the observer) compete with Doppler redshifted regions (receding from the observer). 
The lensed specific flux begins to be greater than the unlensed flux, \ie $M_{\rm X}>1$ at $\theta \gtrsim 70^\circ$.
The model clearly predicts an inclination angle dependence of the magnification ratio $R_{\rm X/O}.$  
Furthermore, the dependence of $R_{\rm X/O}$ on inclination angles is stronger for smaller X-ray disks, larger spins (smaller inner cutoff $r_{\rm ISCO}$), or steeper radial profiles (more weight on the central part close to the black hole).
If the sizes of the X-ray and optical emission regions constrained from microlensing observations are correct, then even the weakest inclination angle dependence among our models, \ie case in which $a=0,$ $r_{\rm disk} = 10\, r_g,$ and flat radial profile,  gives an increase of a factor of two in $R_{\rm X/O}$  when $\theta$ increases from $15^\circ$ to $75^\circ$ (see lower right panel of Figure \ref{fig:Xray_flux_ratio}). 
On the other hand, for the case $a=0.998\,M,$ $r_{\rm disk}=10\,r_g$, and $I_\nu\propto r^{-3},$  $R_{\rm X/O}$ increases by a factor of 10 when $\theta$ increases from $15^\circ$ to $75^\circ.$
For the nearly edge on case, $R_{\rm X/O}$ can increase by another factor of $10$ in extreme cases with small X-ray emission sizes and steep profiles. 

\begin{figure}
	\epsscale{0.7}
	\plotone{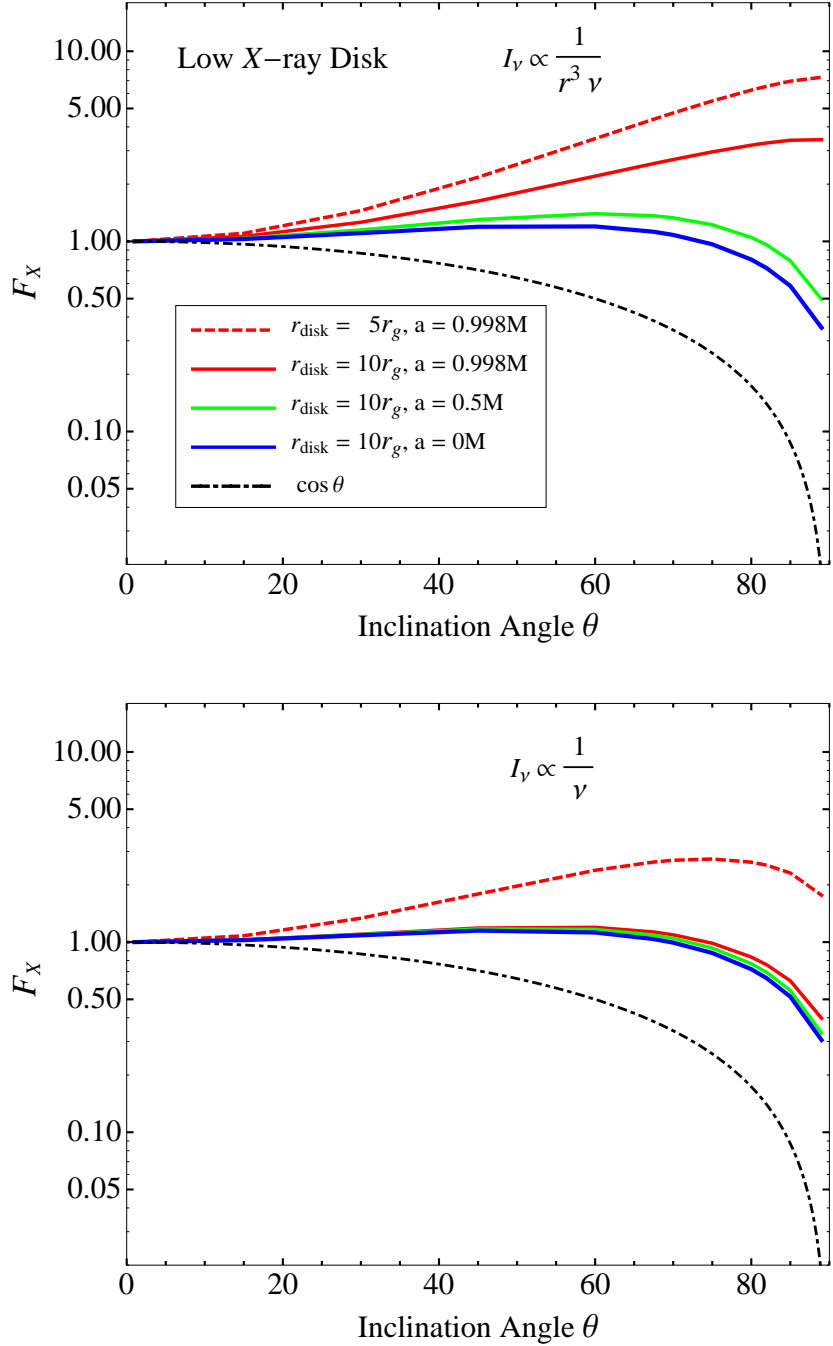}
	\caption{The inclination angle dependence of the observed X-ray flux (normalized to 1 when observed face on) for the low X-ray disk model.
		      For extreme Kerr black hole with steep radial profile (\ie $a=0.998\,M$ and $n=3$, see the red curves in the top panel), the strong lensing magnification overwhelms the demagnification caused by geometrical projection for large inclination angles, and consequently the observed flux increases with $\theta.$
		       	   \label{fig:Xray_flux}} 
\end{figure}

We plot the observed monochromatic flux modulated by strong lensing and geometrical projection as a function of inclination angle $\theta$ in Figure~\ref{fig:Xray_flux}.
For the case of an extreme Kerr black hole case ($a=0.998\,M$) with steep radial profile, the strong lensing effect overwhelms geometrical projection, and consequently,  the observed flux $F_{\nu}$ increases with $\theta.$     
A comparison between the top and bottom panel of Figure~\ref{fig:Xray_flux}  shows that difference between different spins is more significant for the steeper radial profile case (\ie n=3).

\subsection{An X-ray Ball Model}\label{sec:Xball}

\begin{figure}
	\epsscale{0.8}
	\plotone{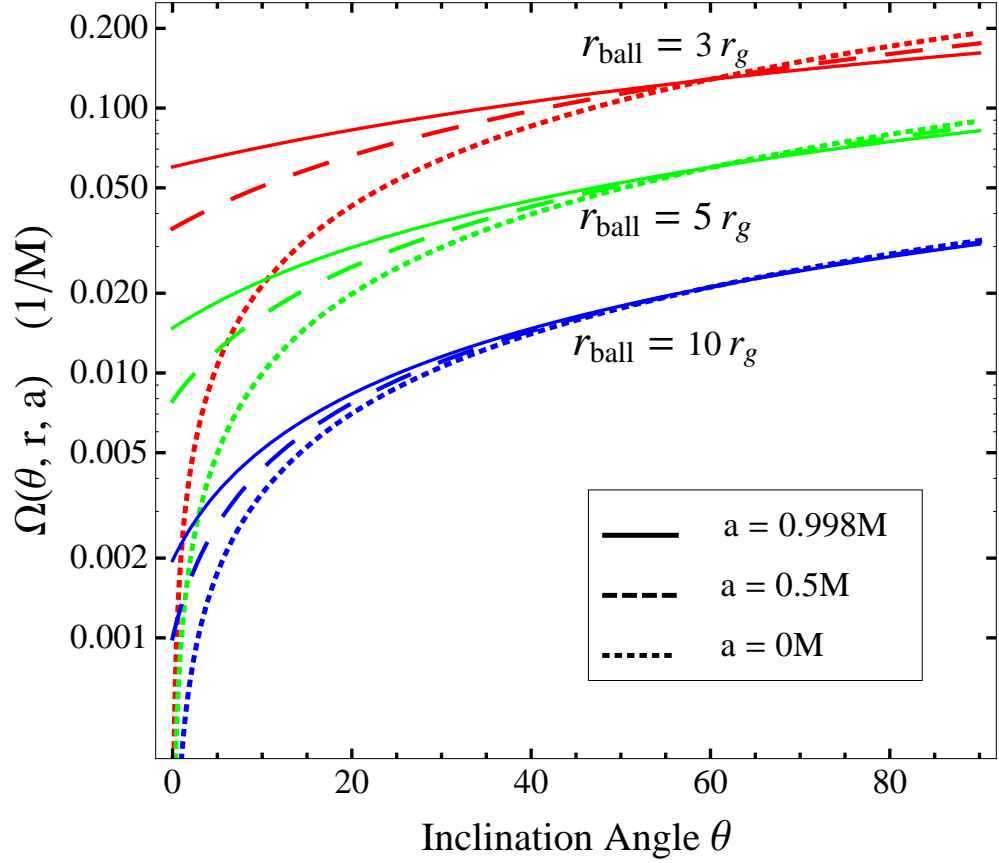}
	\caption{ Angular velocity $\Omega$ as a function of the inclination angle $\theta$ for the X-ray ball model with $r_{\rm ball}=3\,r_g,$ $5\,r_g,$ and $10\,r_g,$ and spin $a=$ 0, 0.5\,M, and 0.998 M.
		 \label{fig:Xball_Omega}} 
\end{figure}

As the second geometrical model for the X-ray emission, we assume that  the X-ray corona is a ball around the central black hole.
Such a corona is often used in the literature (e.g., Schnittman \& Krolik 2010) to study the X-ray polarization of AGN. 
We furthermore assume that the accretion disk is opaque, and integrate over the half sphere  above the accretion disk to obtain the observed X-ray flux. 
We test this X-ray ball model for three radii: $r_{\rm ball}=3\,r_g,$ $5\,r_g$, and $10\,r_g.$ 
For the 4-velocity of the X-ray sphere, we use Eqs.~(\ref{u_K}) and (\ref{RF_Omega}) (Ruszkowski \& Fabian 2000).
We do not consider the radial motion of the X-ray source, and the X-ray sphere is differentially rotating around the polar axis.  
We show $\Omega(\theta)$ for different $r_{\rm ball}$ and spins in Figure~\ref{fig:Xball_Omega}.   
Smaller $r_{\rm ball}$ gives larger $\Omega.$ 
For small to moderate inclination angles, $\Omega$ is dominated by $\omega,$ \ie the frame dragging, and therefore a smaller $a$ gives a smaller $\Omega,$ whereas for a large $\theta$ (near the accretion disk), $\Omega$ is dominated by the Keplerian flow $\Omega_K$, and a smaller $a$ gives a larger $\Omega$ for a prograde flow, see Eq.~(\ref{Omega_K}) and Figure~\ref{fig:Xray_Omega}.   

\begin{figure}
	\epsscale{1.0}
	\plotone{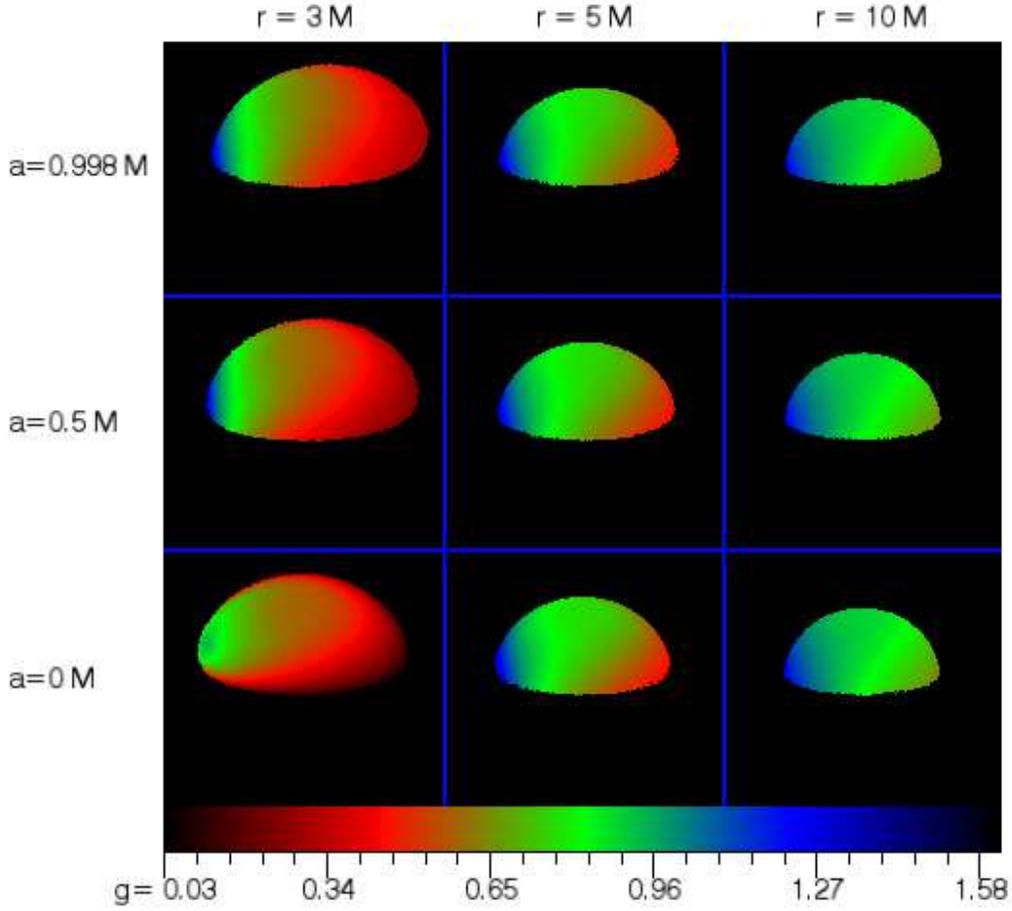}
	\caption{ Lensed and redshifted images for the X-ray ball model.  
	The top, middle, and bottom rows show results for spins $a=0.998\,M,$ 0.5\,M, and 0. The first, second, and last columns are for $r_{\rm ball}= 3\,r_g$, $5\,r_g$, and  $10\,r_g,$ respectively. 
	The observer is at $r_{\rm obs} = 10^6\, r_g$ with an inclination angle $75^\circ.$  
	The area amplification is more significant for  smaller ball radii (a factor $\sim$2.3 for $r_{\rm ball}= 3\,r_g$ case).
	The source on the left hand side is moving toward the observer, and the source on the right hand side is receding from the observer.      
	 \label{fig:Xball_image}} 
\end{figure}

We show the lensing distorted and redshifted image of the (half) X-ray ball above the equatorial plane in Figure~\ref{fig:Xball_image} for an observer at $\theta =75^{\circ},$ for three ball sizes, $r_{\rm ball}= 3\,r_g$, $5\,r_g$, and $10\, r_g$,  and three spins, a = 0, 0.5\,M, and 0.998\,M.  
The images are taken in a way such that the fractional area of the image of the (half) X-ray sphere with respect to  the area of the image would have been the same for each frame,  were there no gravity.
The area amplification is most significant for the $r_{\rm ball}= 3\,r_g$ case (a factor $\sim2.3$ for $\theta=75^{\circ}$) since it is closest to the black hole, and is therefore influenced by the lensing effect most significantly, and least significant for $r_{\rm ball}=10\,r_g$ (about 20\%). 
For the two $a\ne 0$ cases, we observe that the image shape (not the color) is not mirror symmetric with respect to the polar axis, this is most obvious in the $r_{\rm ball}=3\,r_g$ case, because of frame dragging.
The image is stretched more on the right hand side (the receding side).        
Both the gravitational redshift and the Doppler red/blue shift effects are more significant for smaller $r_{\rm ball}$ (see also Figure \ref{fig:Xball_Omega}). 
For example, for the moderate spin case ($a=0.5$), the redshift factor $g\equiv\nu_{\rm o}/\nu_e$ lies between 0.63 and 1.26 for $r_{\rm ball}=10\, r_g$  and between 0.22 and 1.39 for $r_{\rm ball} = 3\, r_g.$ 
 
\begin{figure}
	\epsscale{1.0}
	\plotone{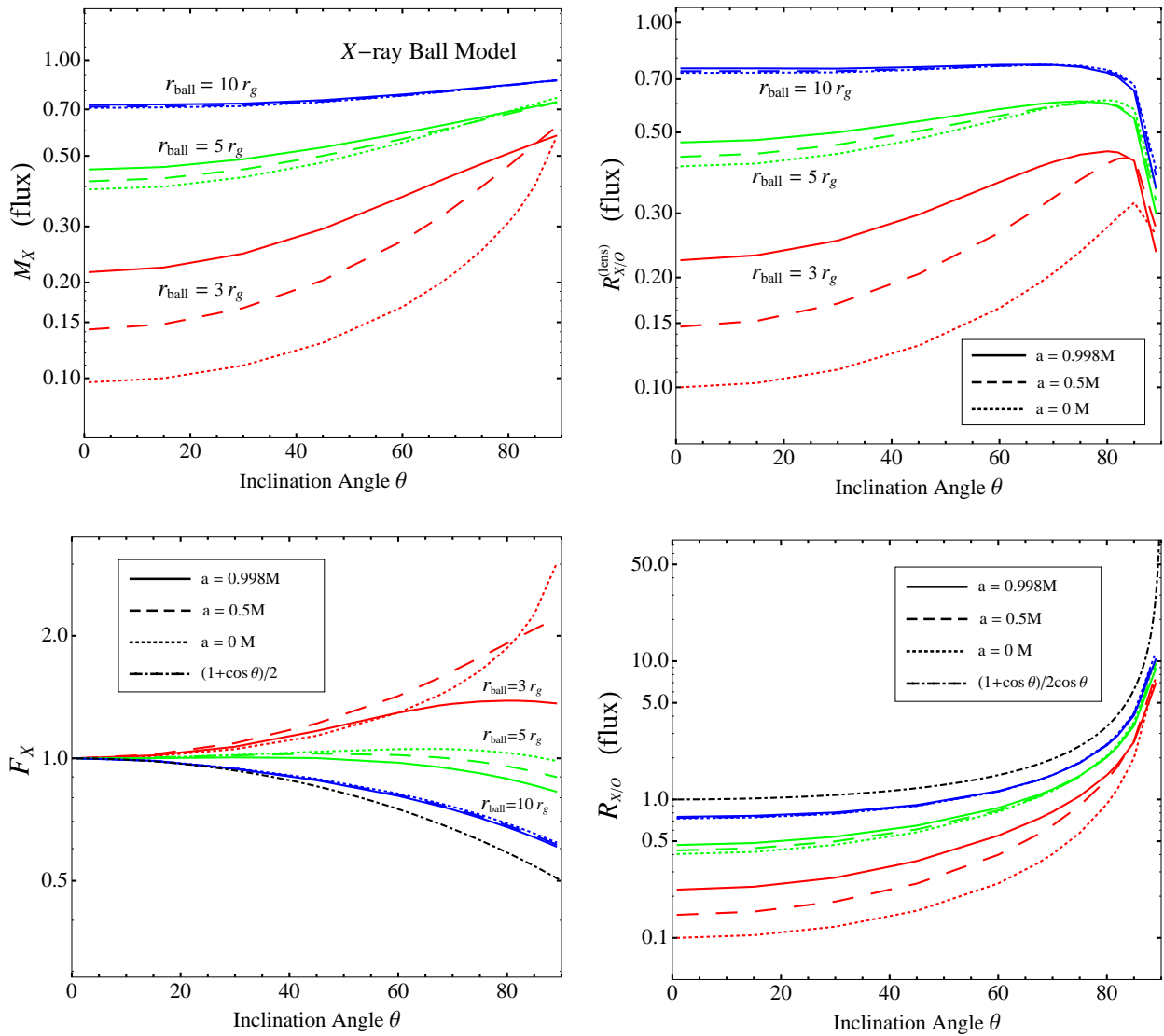}
	\caption{ The inclination angle dependence of the observed X-ray-to-optical flux ratio assuming X-rays emitted from a (half) ball around the black hole with $r_{\rm ball}= 3\,r_g$, $5\,r_g$, and $10\,r_g,$ respectively.
		       We plot the strong lensing magnification of the X-ray flux (the top left panel),  X-ray-to-optical magnification ratio $R_{\rm X/O}^{\rm (lens)}$ (the top right panel),  observed X-ray flux (normalized to 1 when $\theta=0^\circ$, the bottom left panel), and the total correction ratio $R_{\rm X/O}=R_{\rm X/O}^{\rm (lens)}R_{\rm X/O}^{\rm (proj)}$ (the bottom right panel). 
	         \label{fig:Xball_ratio}} 
\end{figure}

For a distant observer  ($r_{\rm obs}\gg 1$) at an inclination angle $\theta,$ the projected image area of a half sphere orthogonal to the line of sight is $\pi r^2(1+\cos\theta)/2$ when the space-time is flat, whereas the lensed image area in Kerr space-time has to be integrated numerically.
Using Eq. (\ref{F_nu}) the monochromatic flux amplification is found to be  
\be
M_{\rm X}
%\equiv \frac{F_{\nu_{\rm o}}}{F^{\rm (flat)}_{\nu_{\rm o}}}
=\frac{\int{g^{\Gamma+2}d\Omega_{\rm o}}}{\int{d\Omega'_{\rm o}}}.
\ee (the $1/r^n$ factor drops out since it is a constant for the ball model, and therefore, the results do not depend on the radial profile).
We plot $M_{\rm X}$ and the lensing magnification ratio $R_{\rm X/O}^{\rm (lens)}$ in Figure \ref{fig:Xball_ratio} (the first row). 
As mentioned earlier, the amplification in the image area is the largest for the $r_{\rm ball}=3\, r_g$ case (Figure \ref{fig:Xball_image}).
However, since the redshift effect dominates over the area amplification, the $r_{\rm ball}= 3\,r_g$ case shows the smallest amplification in the observed flux (Figure \ref{fig:Xball_ratio}, top left panel).  
As for the inclination angle dependence of flux magnification ratio $R_{\rm X/O}^{\rm (lens)}$, we find that the magnification ratio changes by less than 5\% ($15^\circ\lesssim\theta\lesssim 75^\circ$) for all three spins when $r_{\rm ball}=10\, r_g.$  
%Therefore, this X-ray ball model is consistent with the flux-ratio observations. 
For smaller $r_{\rm ball},$ we find that the magnification ratio $R_{\rm X/O}^{\rm (lens)}$ increases by  $\sim$20--40\% for different spins when $r_{\rm ball}= 5\,r_g$  and by  $\sim$80--130\% for $r_{\rm ball}= 3\, r_g$ with different spins.
For large inclination angles, the magnification ratio $R_{\rm X/O}^{\rm (lens)}$ decreases with $\theta$ which is different from the X-ray disk model in the previous subsection (see Figures \ref{fig:Xray_Area_ratio} and \ref{fig:Xray_flux_ratio}), and is mainly caused by the fact that in the case of no lensing the projected area of the thin optical disk approaches zero when $\theta$ approaches $\pi/2$ (the amplification of image area therefore approaches infinity) whereas that of the (half) X-ray ball approaches  $\pi r^2/2$ (the area magnification is therefore finite). 

Since the geometry of the X-ray ball model is different from that of the optical emission (a thin disk), the projection ratio $R_{\rm X/O}^{\rm (proj)}=(1+\cos\theta)/2\cos\theta$ is not a constant.
Consequently both Kerr strong lensing and the projection effect contribute to the inclination angle dependence of the observed X-ray-to-optical flux ratio, and $R_{\rm X/O}\ne R_{\rm X/O}^{\rm (lens)}$ (see Eq.~\ref{RXO}).
We plot $R_{\rm X/O}$ in Figure~\ref{fig:Xball_ratio} (the bottom right panel).
$R_{\rm X/O}$ can increase by a factor of $\sim$10 for large inclination angle ($\theta \gtrsim 70^\circ$) where the major contribution comes from the geometrical projection instead of general relativity. 
We also show the observed flux as a function of $\theta$ in Figure~\ref{fig:Xball_ratio} (the bottom left panel).
Similar to the low X-ray disk model, $F_{\rm X}$ can increase with $\theta$ for small ball radii.

%This is very interesting, because it was thought by the community that the X-ray-to-optical flux ratio is the same for normal quasars and broad absorption line (BAL) quasars after correcting for absorptions. 
%The X-ray disk model in the previous subsection gives monotonically increasing $R_{\rm X/O}$ (from less than 1 to greater than 1, see Figure \ref{fig:Xray_flux_ratio}).  
%For BAL this implies that the X-ray emission is boosted by the strong gravity compared with the optical emission, and therefore the absorption in X-rays was under-estimated, whereas the X-ray ball model in this section gives  $R_{\rm X/O}<1$ for BALs which is exactly the opposite.

%We show in Figure \ref{fig:Xball_PDF} the p.d.f.  $P_{R_{\rm X/O}}$ of the X-ray-to-optical magnification ratio for one  X-ray ball model where $r_{\rm ball}= 3\,r_g.$ 
%Similar to the X-ray disk model in the previous section, the p.d.f is non-uniform, and skewed (it has greater value where the $R_{\rm X/O}(\theta)$ curve is flat, see Eq.\,(\ref{pdf})).  

%---------------------------------------------------
\subsection{High X-Ray Disk Model}

\begin{figure}
	\epsscale{0.8}
	\plotone{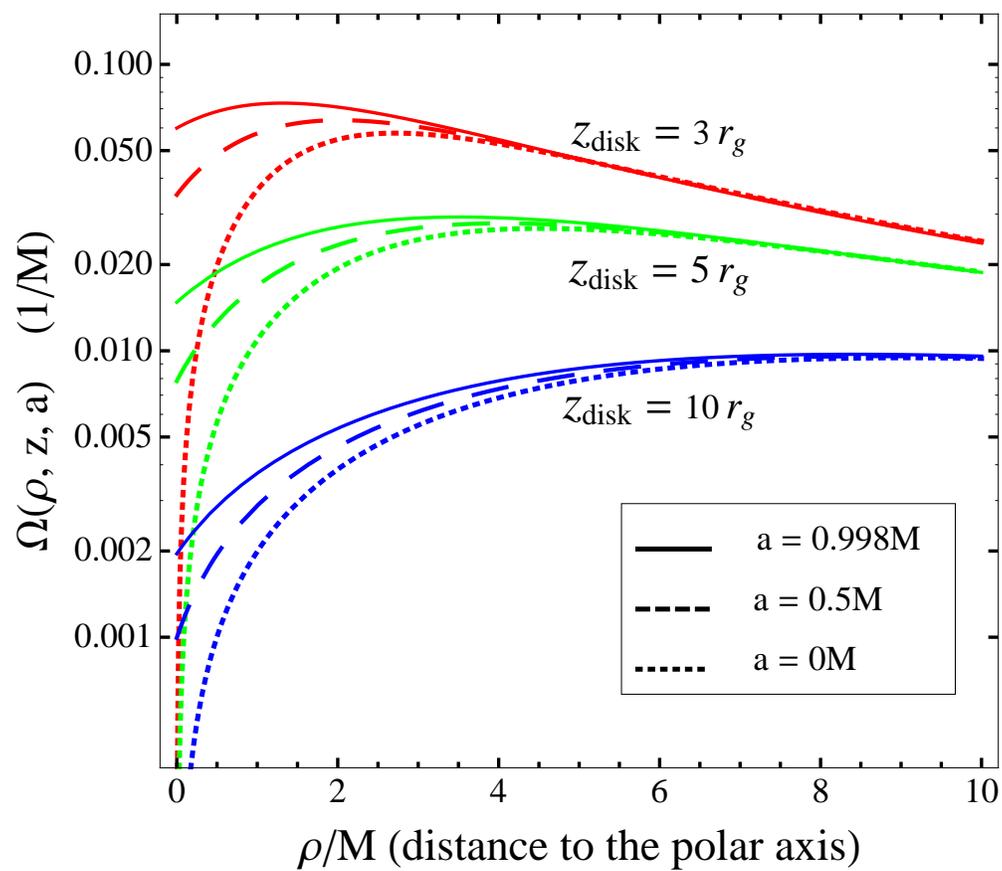}
	\caption{ Angular velocity $\Omega$ as a function of $\rho,$ distance to the polar axis,  for the high X-ray disk model with disk height $z_{\rm disk} = 3\,r_g,$ $5\,r_g,$ and $10\,r_g,$ and spin $a=$ 0, 0.5\,M, and 0.998 M. 
	         \label{fig:Xover_Omega}} 
\end{figure}

As the last example of a X-ray emission model, we consider a thin X-ray disk of radius $10\,r_g$ which is above the black hole (parallel to the equatorial plane) with height $z= 3\,r_g$, $5\,r_g$, and $10\,r_g$, respectively. 
This is motivated by the light bending model of Fabian \& Vaughan (2003) where the authors tried to explain the rapid variability in the continuum observed for MCG6$-$30$-$15 using the fact that gravitational light bending by the strong field of the central black hole amplifies small changes in the X-ray source height (see also Miniutti et al.\ 2003; Miniutti \& Fabian 2004).  
Intuitively, a larger $z$ means greater distance of the X-ray disk from the black hole, and therefore less influence from the strong gravitational field, and this might give smaller inclination angle dependence in $R_{\rm X/O}$ when $\theta$ is not too large. 
We plot the angular velocity profile in Figure~\ref{fig:Xover_Omega}. 
From Figure~\ref{fig:Xover_Omega}, the disk with the smallest height (closest to the black hole) has the largest velocity (for the same $\rho,$ distance to the polar axis, and spin parameter a). 

\begin{figure}
	\epsscale{1.0}
	\plotone{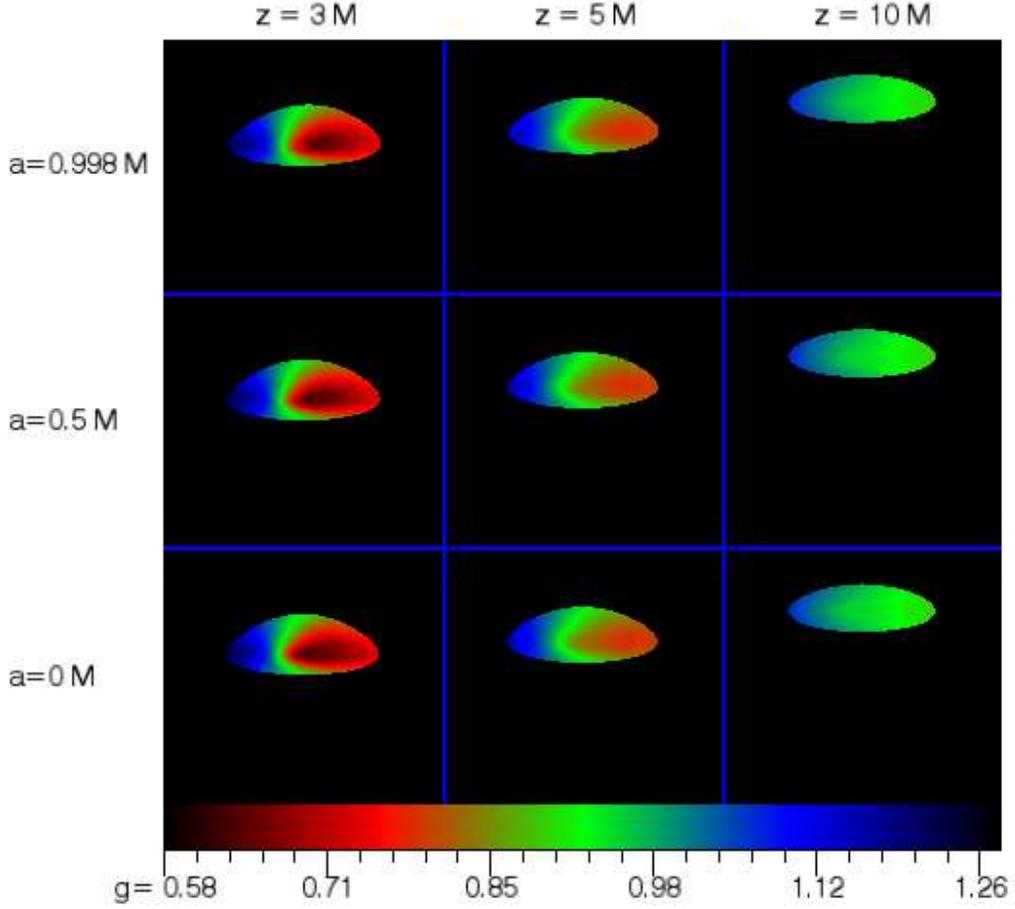}
	\caption{  Lensed and redshifted images for the high X-ray disk model. 
	The disk size is $r_{\rm disk}=10\, r_g$.  
	The top, middle, and bottom rows show results for spins $a=0.998\,M,$ $0.5\,M$, and 0. 
	The first, second, and last columns are for height $z_{\rm disk}= 3\,r_g$, $5\,r_g$, and $10\,r_g,$ respectively. 
	The observer is at $r_{\rm obs} = 10^6\, r_g$ with an inclination angle $75^\circ.$ 
	The source on the left hand side is moving toward the observer, and the source on the right hand side is receding from the observer.      
	 \label{fig:Xover_image}} 
\end{figure}

We show images of the lensed disk above the black hole as seen by an observer at $\theta=75^{\circ}$ in Figure \ref{fig:Xover_image} for three heights, and three spins.
The area amplification is most significant for the height $z=3\,r_g$ case (a factor of  $\sim$$1.9$ for all three spins), and least significant for the $z=10\,r_g$ case (a factor about $1.5$ for all three spins). 
The redshift effect (gravitational and Doppler) is also most significant for the $z= 3\, r_g$ case (see the first column of Figure~\ref{fig:Xover_image}). 
For example, for moderate spin $a=0.5,$ the $g$ factor lies between 0.84 and 1.02 for $z_{\rm disk}=10\, r_g$ case, and between 0.58 and 1.16 for $z_{\rm disk}= 3\,r_g$ case. 
Since points in the  interior of the disk are closer to the black hole, they are more influenced by gravitational redshift.  
Since the  $\Omega(\rho)$ profile (Ruszkowski \& Fabian 2000) is not monotonic for this model (note that $\rho$ increases with $\theta$ for fixed $z_{\rm disk}$), %and the linear velocity profile, $\rho*\Omega(\rho),$ is not rapidly increasing with $\rho$ for $\rho \gtrsim 4\,r_g,$
the points on the disk with most significant Doppler blue/redshift can lie in the interior of the disk for small $z_{\rm disk}$. 
For example, when $z_{\rm disk}= 3\, r_g,$ the largest redshift occurs in the interior of the disk instead of on the boundary (Figure \ref{fig:Xover_image}, the first column).
  
\begin{figure}
	\epsscale{1.0}
	\plotone{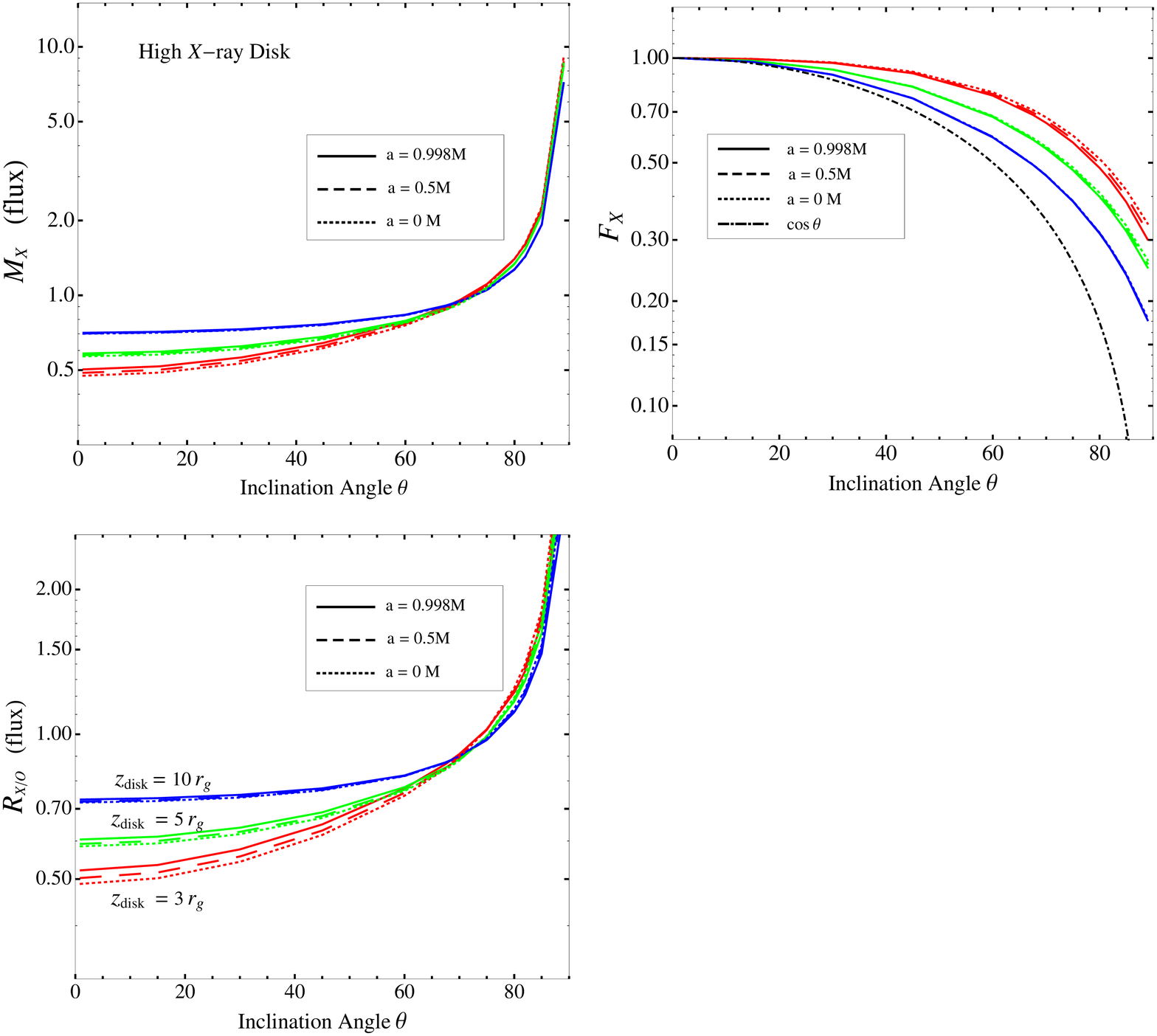}
	\caption{ Inclination angle dependence of the X-ray-to-optical flux ratio for the high X-ray disk model.  
	We assume that X-rays are emitted from a disk with $r_{\rm disk}= 10\, r_g,$ above the black hole, with height $z= 3\,r_g$, $5\,r_g$, and $10\,r_g$, respectively.
	We plot the lensing magnification of the X-ray flux $M_{\rm X}$ (the top left panel), the observed flux (normalized to 1 when observed face on, the top right panel), and the X-ray-to-optical magnification ratio $R_{\rm X/O}=R_{\rm X/O}^{\rm (lens)}$ (the bottom panel) as function of inclination angle $\theta.$
          \label{fig:Xover_ratio}} 
\end{figure}

We choose no cutoff at the center of the disk, and use the radius-independent intensity profile (since the smallest disk height we used is $3\,r_g$, the dependence of the results on the radial profile should  be much weaker than the low X-ray disk model in \textsection\ref{sec:Xray} where the case most sensitive to the radial profile is when $a=0.998$ and $r_{\rm ISCO}=1.23\, r_g$). 
This gives less weight to the central part of the disk which is closest to the black hole compared with the $n=3$ ($I_\nu \propto r^{-3}$) case, and will yield more conservative constraints.
The projection effect (\ie the $\cos\theta$ factor) is the same for X-ray and optical emission, consequently $R_{\rm X/O}=R_{\rm X/O}^{\rm (lens)}$.
We show the magnification ratio $R_{\rm X/O}$  in Figure~\ref{fig:Xover_ratio}.
Since the black hole spin contributes to the space-time curvature only in higher orders, and the smallest X-ray  disk height $z$ we test is $3\, r_g$, we found that the results for different spins do not differ significantly.  
For $15^\circ \lesssim \theta \lesssim 75^\circ,$ the magnification ratio  $R_{\rm X/O}$ (flux)  increases by about 85\%, 55\%, and 27\% for $z= 3\,r_g$, $5\,r_g$, and $10\,r_g$ respectively. 
Therefore, a thin X-ray disk of radius $10\,r_g$ with Ruszkowski \& Fabian (2000) velocity profile is consistent with observed $\alpha_{\rm ox}$ for normal quasars even when it is   above the central black hole with a height as low as $3\, r_g$ ($R_{\rm X/O}\sim2$ corresponds to $\Delta \alpha_{\rm ox}\sim 0.1,$ well within the observed scatter of $\alpha_{\rm ox}$, see next section).
%We plot  the p.d.f. function $P_{R_{\rm X/O}}$ in Figure \ref{fig:Xover_PDF}. 
%We show results for three disk heights, but only for one spin,  $a=0.998\,M$ (the results of the other two spins are very similar, see Figure \ref{fig:Xover_ratio}).
%We see from Figure \ref{fig:Xover_PDF} that the p.d.f function is steep, nonuniform, and strongly skewed.
For high inclination angles (corresponding to BALs), the magnification ratio $R_{\rm X/O}$ in observed monochromatic flux is larger than one (say, a  factor of 2, see Figure  \ref{fig:Xover_ratio}) which is similar to the low X-ray disk model (although less significant). %, since it is above the black hole with some minimum height, instead of embedded in the equatorial plane). 
This implies that the absorption in the X-ray band might be under-estimated. %, contrary to the X-ray ball model in which $R_{\rm X/O}$ is less than 1 for high inclination angles.  

In the so-called light-bending model (Miniutti et al.\  2003; Fabian \& Vaughan 2003; and Miniutti \& Fabian 2004) the authors used the variation of the X-ray source height in a strong gravitational field as one interpretation for the observed large variability of the X-ray power-law component compared to that of the reflection component (in particular the fluorescent iron line emission). 
Choosing a ring-like source with distance $\rho_s = 2\,r_g$ from the z-axis and assuming that the source is co-rotating with the accretion disk with the same orbital velocity as the underlying disk, Miniutti \& Fabian (2004) found that the observed flux from the (direct) power-law component can increase by a factor of $\sim$20 when the ring hight $z$ increases from $1\,r_g$ to $20\,r_g,$ or a factor of  $\sim$5 when $z$ increases from $3\,r_g$ to $10\,r_g$ (for inclination angle $\theta = 30^\circ$).  
For our disk model with $r_{\rm disk} = 10\,r_g$ and Ruszkowski \& Fabian (2000) velocity profile, we found that  if  such a X-ray disk is  observed with an inclination angle $\theta = 30^\circ$, the observed X-ray flux can increase by a factor of $\sim$40\% when the disk height changes from $3\,r_g$ to $10\,r_g$  (see Figure~\ref{fig:Xover_ratio}). 
Our results are qualitatively consistent with Muniutii \& Fabian (2004). 
The quantitative difference, \ie our models shows much smaller variation in observed flux of the power-law component, is not surprising,  since we have assumed a very different geometry for the source emission (a disk of size $10\,r_g$ versus a ring with radius $2\, r_g$).

%-----------------------------------------------------------------------------
\section{Comparison With Observations}\label{sec:observation}

An important parameter characterizing the flux ratio between X-ray and optical emission is $\alpha_{\rm ox}$ (Tananbaum et al.\ 1979)
\be
%\alpha_{\rm ox}\equiv \log \frac{ F^{\rm X}_{\nu_{\rm o}}}{ F^{\rm O}_{\nu'_{\rm o}}}\Bigg/ \log\frac{\nu_{\rm o}}{\nu'_{\rm o}} = 0.3838 \log \frac{F_{\rm 2\, keV}}{F_{\rm 2500\, { \AA}}}.
\alpha_{\rm ox}\equiv 0.3838 \log \frac{F_{\rm 2\, keV}}{F_{\rm 2500\, { \AA}}}.
\ee 
Some authors use $\rm 3000\,\AA$ for the optical emission with the constant in the front as 0.372, and this does not change the conclusion of this paper.
There is evidence showing that the observed $\alpha_{\rm ox}$ in quasars are strongly anti-correlated with quasar luminosities in both the optical and X-ray bands; however, no significant redshift evolution was found (Strateva et al.\ 2005; Steffen et al.\ 2006; Lusso et al.\ 2010).
The intrinsic scatter of $\alpha_{\rm ox}$ is about 0.2--0.3 at a given luminosity (Steffen et al.\ 2006;  Lusso et al.\ 2010).
There are several factors contributing to this scatter including quasar variability and intrinsic differences between quasars.

In this paper, we have investigated variations in $R_{\rm X/O}$ due to lensing and projection effects which are functions of the inclination angle $\theta$ (see Figures~\ref{fig:Xray_flux_ratio}, \ref{fig:Xball_ratio}, and \ref{fig:Xover_ratio}), and therefore, the orientation effect also contributes to the distribution of $\alpha_{\rm ox}$.
To test the significance of strong lensing effects on the observed X-ray-to-optical flux ratios, we separate the contributions to \aox\ as,
\be\label{ox:modulation}
\alpha_{\rm ox}= \alpha^{\rm intr}_{\rm ox}+0.3838\log R_{\rm X/O},
\ee 
where $R_{\rm X/O}$ includes the orientation induced contribution and $\alpha^{\rm intr}_{\rm ox}$ represents other contributions such as spectral variability and intrinsic differences between quasars.
 We assume the intrinsic distribution of $\alpha^{\rm intr}_{\rm ox}$ to be Gaussian with mean value $\mu=-1.5$ and  scatter $\sigma=0.2$.
Assuming a uniform distribution for $\alpha^{\rm intr}_{\rm ox}$, e.g., between $(-2,\,-1),$ does not change our conclusions. 
The distributions of $R_{\rm X/O}$ are calculated for the three X-ray models which we have considered,  assuming observers are randomly distributed in all solid angles, i.e.,  a uniform distribution in the variable $\mu\equiv \cos\theta$ ($0<\mu<1$).
We use Monte-Carlo simulations to combine the two distributions to obtain the distributions of \aox\ using 500,000 realizations.

%, the dependence of  $R_{\rm X/O}$ on $\theta$ makes $R_{\rm X/O}$ a random variable with probability distribution function (p.d.f.)
%\be\label{pdf}
%P_{R_{\rm X/O}}(R_{\rm X/O})=\frac{\sin\theta}{R'_{\rm X/O}(\theta)}.
%\ee 
%If the magnification ratio $R_{\rm X/O}$ indeed depends on the inclination angle $\theta$, it will be a random variable with nontrivial distribution determined by black hole physics and corona geometry.
%Therefore, the observed distribution of $\alpha_{\rm ox}$ will be modulated by that of $R_{\rm X/O},$ \ie
%,  and compute the p.d.f. of lensed $\alpha_{\rm ox}$ for the three X-ray models considered in this paper.

We assume that differences between normal quasars, BALs, and obscured quasars are pure observational effects caused by different inclination angles.  
We furthermore assume that obscured quasars are about  $20\%$ of the total population (\ie Type I + Type II quasars, Hasinger 2008) since we are focusing on the high luminosity regime,  and BALs count for 40\% of Type I quasars (Dai et al.\ 2008; Shankar et al.\ 2008).
Although the 20\% Type II fraction might be lower than some other measurements, here we use this as a fiducial number to test the predictions of our model.
In addition, some BALs such as LoBALs and FeLoBALs may have Compton thick absorptions, which can be classified as Type II quasars.  
The fraction of these objects is non-negligible, 4--7\% (Dai et al.\ 2012), which can compensate for our lower assumed Type II fraction.
Consequently,  we assume $0^\circ < \theta^{\rm Normal}\le 58.7^\circ <\theta^{\rm BAL}\le 78.5^\circ.$
We do not calculate the observed \aox\ distributions for Type II quasars because their $\alpha_{\rm ox}$ is hard to study due to the heavy Compton-thick absorption.
This also avoids very large inclination angles where the thin disk approximation assumed for optical emission and the two X-ray disk models can be inaccurate (we have assumed zero thickness for the thin disk models; however, the finite thickness of the disks must be considered for nearly edge on cases when computing unlensed quantities such as image area).
The inclination dependence of $R_{\rm X/O}$ is the same as the strong lensing $R_{\rm X/O}^{\rm (lens)}$ for the low and high X-ray disk models, because the projection factor $\cos{\theta}$ is canceled out considering both the optical and X-ray emission.
For the X-ray ball model, the projection effect, $R_{\rm X/O}^{\rm (proj)} = (1+\cos\theta)/2\cos\theta$, is not canceled out, and both lensing and projection effects contribute to the inclination dependent flux ratios.
%  there is another source of scatter of $\alpha_{\rm ox}$ caused by randomly distributed inclination angles: the different geometries assumed for optical and X-ray emission (thin disk for optical emission, and half sphere for X-rays) will make X-ray-to-optical flux ratios depend on inclination angles because of the projection effect ($\propto (1+\cos\theta)/\cos\theta $) even for flat space-time (there is no such scatter for the other two X-ray models).
%We have taken this effect into account when computing the observed $\alpha_{\rm ox}$ for the X-ray ball model.        

\begin{figure}
 \epsscale{1.0}
        \plotone{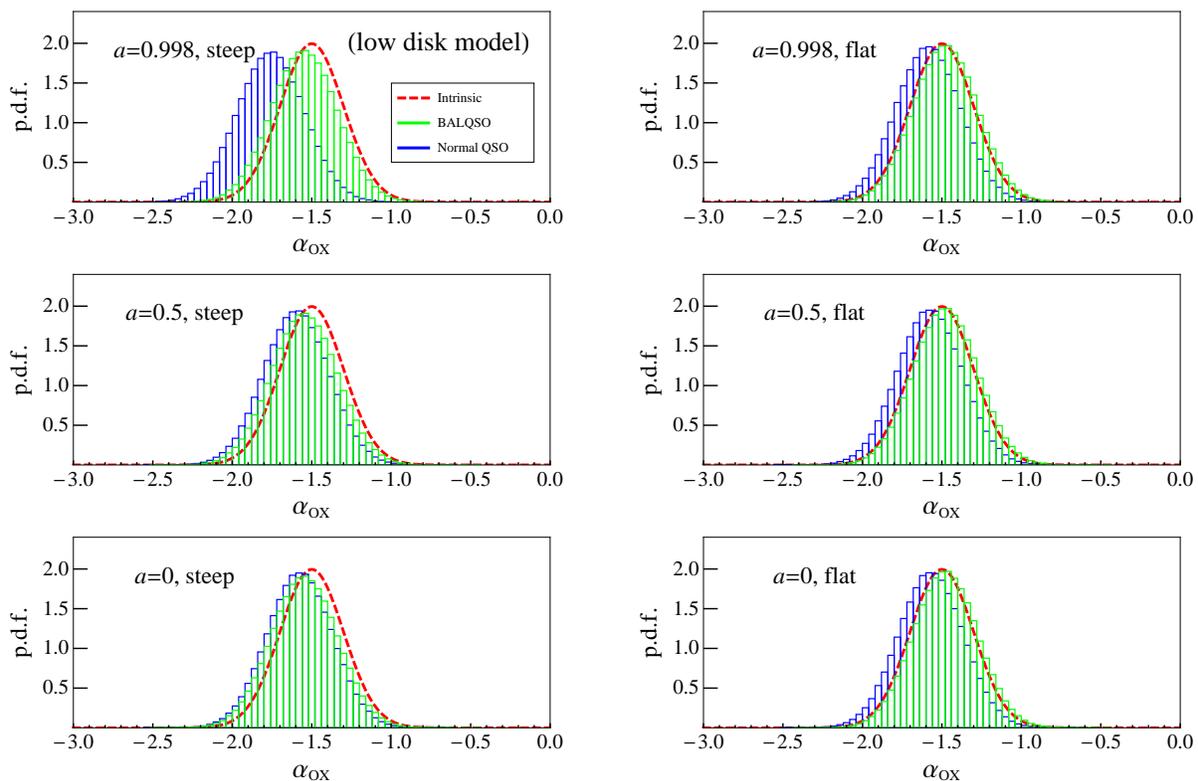}
        \caption{ Histograms of $\alpha_{\rm ox}$ for normal (blue) and BAL quasars (green) including the Kerr strong lensing effects for the low X-ray disk model.
                        The intrinsic scatter of $\alpha_{\rm ox}$ is assumed to be Gaussian with $\mu=-1.5,$ $\sigma=0.2$ (dashed red line).  
                        $\langle\alpha_{\rm ox}\rangle^{\rm Normal}$ is smaller than $\langle\alpha_{\rm ox}\rangle^{\rm BAL}$ by $\sim$0.1--0.2.
                        See also Table~\ref{tab:Gauss}.
                        \label{fig:Xray_PDF_new} }
\end{figure}

\begin{figure}
 \epsscale{1.0}
        \plotone{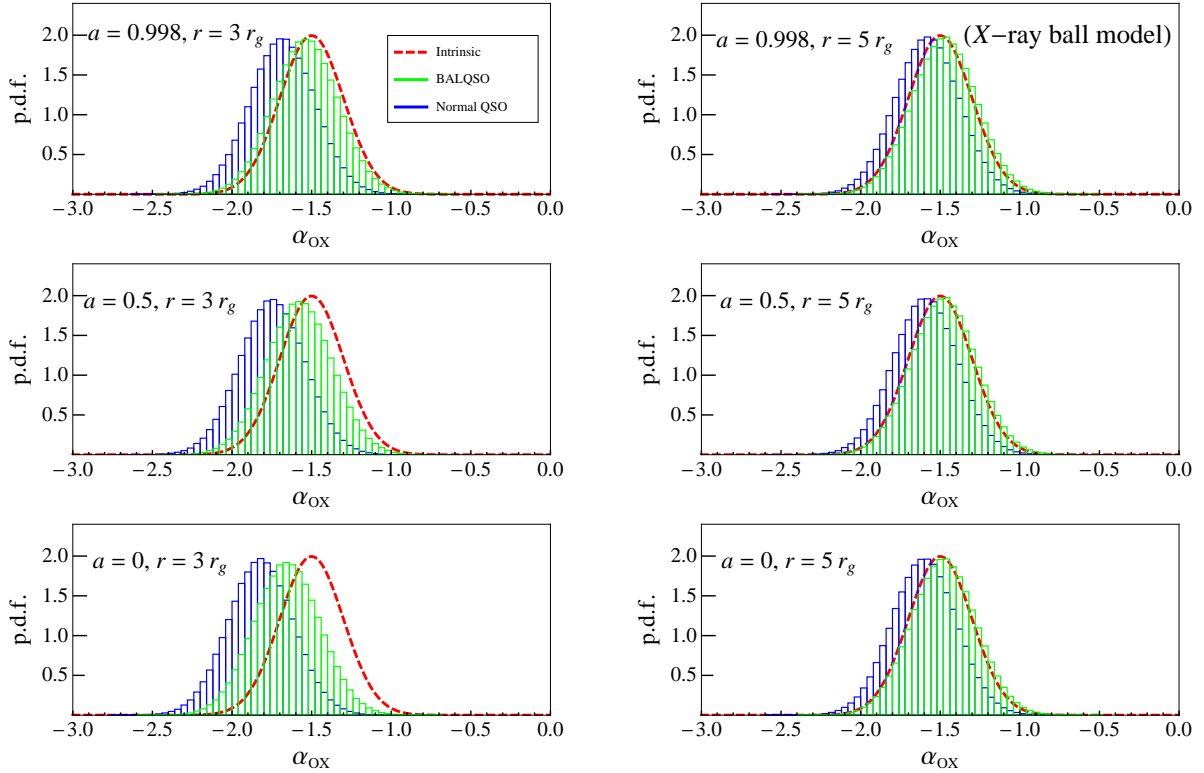}
        \caption{Histograms of $\alpha_{\rm ox}$ for normal (blue) and BAL quasars (green) including the Kerr strong lensing and projection effects for the X-ray ball model, 
                        where the extra projection effect is caused by the different geometries of X-ray and optical emission.                      
                        The intrinsic scatter of $\alpha_{\rm ox}$ is assumed to be Gaussian with $\mu=-1.5,$ $\sigma=0.2$ (dashed red line).  
		      $\langle\alpha_{\rm ox}\rangle$ for normal quasars is smaller than that of BALs by $\sim$0.15 and  $\sim$0.11 for $r_{\rm ball}= 3\,r_g,$ and $5\,r_g,$ respectively.  
		     The contribution of the projection effects to $\Delta\langle\alpha_{\rm ox}\rangle$ is $\sim$0.08, independent of the size of the X-ray ball.                     
                        See also Table~\ref{tab:Gauss}.
                        \label{fig:Xball_PDF_new} }
\end{figure}

\begin{figure}
 \epsscale{0.8}
        \plotone{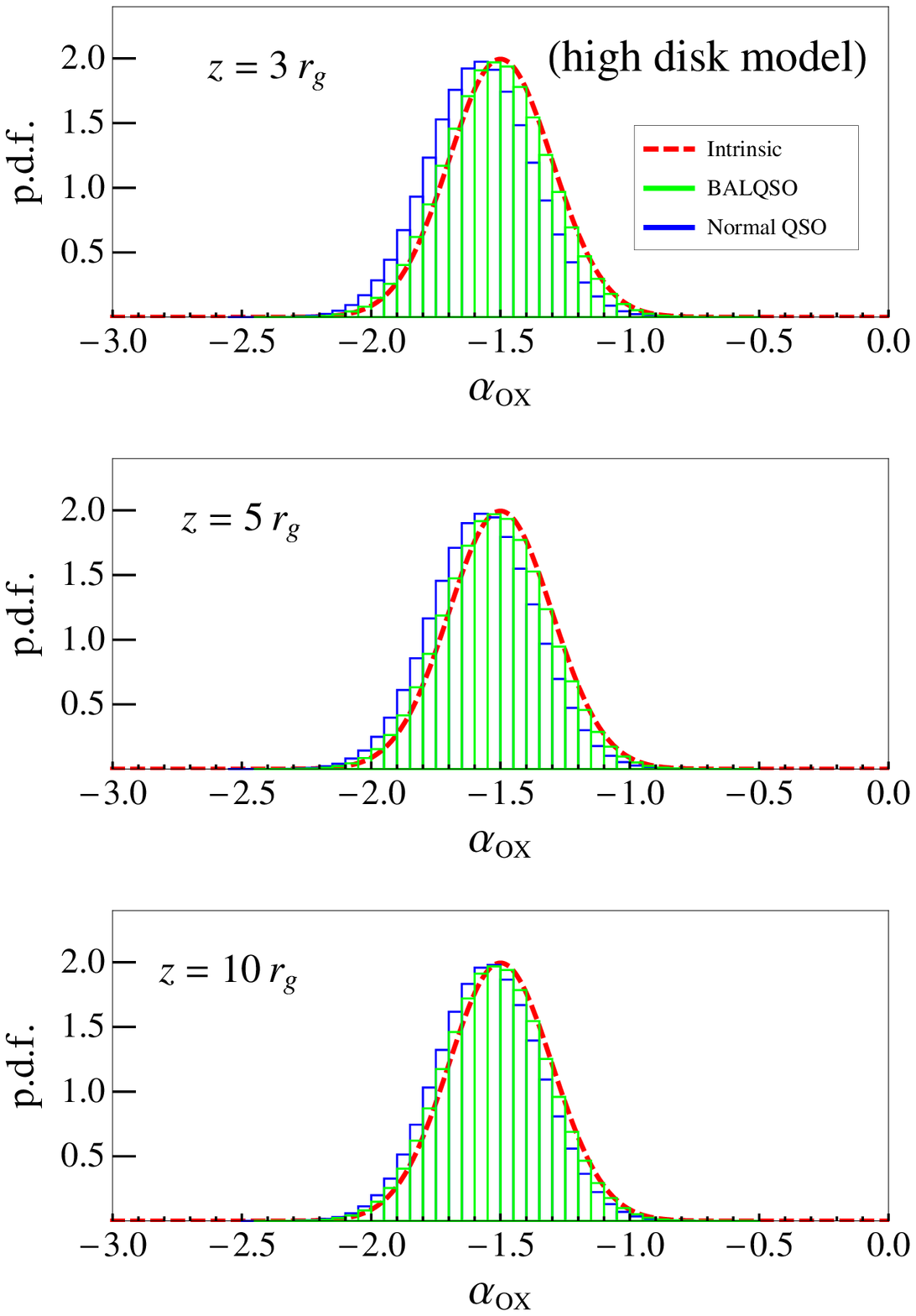}
        \caption{ Histograms of $\alpha_{\rm ox}$ for normal (blue) and BAL quasars (green) including the Kerr strong lensing effects for the high X-ray disk model.
                        The intrinsic scatter  $\alpha_{\rm ox}$ is assumed to be Gaussian with $\mu=-1.5,$ $\sigma=0.2$ (dashed red line).  
                        We find  $\langle\alpha_{\rm ox}\rangle^{\rm BAL}-\langle\alpha_{\rm ox}\rangle^{\rm Normal}\le0.06$.
                        See also Table~\ref{tab:Gauss}.                   
                             \label{fig:Xover_PDF_new} }
\end{figure}

The results for the three X-ray models are shown in Table~\ref{tab:Gauss} and Figures~\ref{fig:Xray_PDF_new}--\ref{fig:Xover_PDF_new}. 
For each X-ray model, we generate the distribution of the observed $\alpha_{\rm ox},$ and computed its mean, standard deviation, and skewness,  \ie $\langle\alpha_{\rm ox}\rangle,$ $\sigma_{\alpha_{\rm ox}},$ and $\cal S,$ respectively, through Monte-Carlo simulation.
The extra scatter added to $\alpha_{\rm ox}$ by the Kerr strong lensing (also by different geometries for the X-ray ball model) is insignificant. 
We have assumed $\alpha^{\rm intr}_{\rm ox}$ to be Gaussian with $\sigma=0.2$ whereas the largest $\sigma$ in the observed $\alpha_{\rm ox}$ is less than 0.24 (corresponding to $\sigma^{\rm (lens)}\le 0.13$) among all models considered in Table~\ref{tab:Gauss}.
Because the scatter of \aox\ is constrained in normal quasars (non-BALs, $0^\circ < \theta^{\rm Normal}\le 58.7^\circ$ in our model), the differential lensing amplification between optical and X-ray emission is small at these small inclination angles.  Therefore, to match the observed scatter in \aox, 0.2--0.3, we have to assume most of the scatter comes from other factors not related to the orientation effect, such as spectral variability or intrinsic differences between quasars.
The skewness of the simulated $\alpha_{\rm ox}$ is also small with the largest case about 0.03.
This is not surprising since distribution of $\alpha_{\rm ox}$ is the convolution of $\alpha^{\rm intr}_{\rm ox}$ with zero skewness and $R_{\rm X/O}$, where we assumed a larger scatter in $\alpha^{\rm intr}_{\rm ox}$. 
%Among the X-ray models considered in this paper, only the low X-ray disk model with extreme spin and steep radial profile shows skewness which might be observationally tested in future, \ie the skewness ${\cal S}$ is about 0.15 for all quasars (normal+BAL), and about 0.08 for BALs.
If the measured intrinsic scatter of \aox\ in normal quasars were smaller, the skewness of the distribution would be a powerful tool to constrain corona geometry.
Given the large measured scatter in non-BALs, which requires a larger  $\alpha^{\rm intr}_{\rm ox}$, it is probably not realistic to try to discriminate corona geometries based on the extra scatter or skewness of the lensed distribution of $\alpha_{\rm ox}.$  

Nevertheless, we find that the mean of $\alpha_{\rm ox},$ $\langle\alpha_{\rm ox}\rangle,$ can be shifted significantly from the intrinsic values. 
For example, for the low X-ray disk model with the extreme spin and the steep radial profile, $\langle\alpha_{\rm ox}\rangle^{\rm Normal}$ is smaller than $\langle\alpha_{\rm ox}\rangle^{\rm intrinsic}$ (equal to $-1.5$) by $\sim$0.25. 
The difference in $\alpha_{\rm ox}$ between normal and BAL quasars (see the last column of Table~\ref{tab:Gauss}) can be as large as $\sim$0.2.
For the low X-ray disk and the X-ray ball model, the difference in $\langle\alpha_{\rm ox}\rangle$ between normal and BAL quasars is $\ge 0.1$.
For the high X-ray disk model, the splitting of $\langle\alpha_{\rm ox}\rangle$ between BALs and non-BALs is $\le 0.06$.
Therefore, comparing large samples of observed, absorption corrected \aox\ between BALs and non-BALs, it may be possible to differentiate corona geometry using the orientation effects.
Here, the absorptions in BALs should be determined from spectral properties without assuming the same SEDs for BALs and non-BALs.
%Another interesting result is that for the high/low X-ray disk model, $\alpha_{\rm ox}$ is shifted toward opposite direction for normal quasars and BALs, whereas for the X-ray ball model, $\langle\alpha_{\rm ox}\rangle$ is smaller than the intrinsic value for both normal and BAL quasars, see Figures~\ref{fig:Xray_PDF_new}--\ref{fig:Xover_PDF_new}. 

A number of studies show that BALs and normal quasars have the same intrinsic spectral energy distribution, if the observed soft X-ray weakness of BALs is caused by intrinsic absorption with column densities $\sim$$10^{22}$--$10^{24}~\cmsq$ (e.g., Green et al.\ 2001; Gallagher et al.\ 2002).
For all three X-ray models considered in this paper, we found $\langle\alpha_{\rm ox}\rangle^{\rm BAL}>\langle\alpha_{\rm ox}\rangle^{\rm normal}$ (Table~\ref{tab:Gauss}). 
This implies that on average X-ray emission is boosted more for BALs than for normal quasars.
Therefore, the aforementioned high column densities might still be an underestimate of the intrinsic X-ray absorption of BALs.   
For example, let us consider a BAL quasar at $z=2$ with $\nh=2\times10^{23}~\cmsq$ determined to correct the \aox\ of this BAL to be the same as non-BALs. 
If $R^{\rm BAL}_{\rm X/O}/R^{\rm normal}_{\rm X/O}=2$ between a BAL and a normal quasar based on our model calculations, corresponding to $\Delta \alpha_{\rm ox}\sim0.11$ and a typical value based on Table~\ref{tab:Gauss}, then an absorption of column density $\sim$$10^{24}~\cmsq$ is needed for the BAL after including the strong lensing boost of the X-ray emission for BALs.
Even the smallest shift in $\langle\alpha_{\rm ox}\rangle$ between BAL and normal quasars, \ie 0.02 (corresponds to $R_{\rm X/O}\sim$1.13, see Table~\ref{tab:Gauss}), can cause an under-estimate of the intrinsic absorption by $\Delta \nh = 2\times 10^{23}~\cmsq$.
It is therefore important to consider the effects of the Kerr strong gravity in order to accurately measure the X-ray absorptions in BALs.

%Type I (normal) quasars are observed with an inclination angle $10^\circ\lesssim \theta \lesssim 75^\circ,$ whereas broad absorption line quasars (BALs) are observed with large inclination angles ($\theta \gtrsim 75^\circ$).
%Steffen et al.\ (2006) constrain the scatter of UV-to-X-ray flux ratios to be within a factor of $\sim$10 using a sample of 333 optically selected AGN from the Sloan Digital Sky Survey, COMBO-17 Survey, and Bright Quasar Survey (see also Strateva et al.\ 2005). 

%-------------------------------------------------------------------------------
\section{Conclusion}\label{sec:conclusion}

We have developed a ray-tracing code for the Kerr metric based on numerical integrations of the $1^{\rm st}$ order Hamilton equations for massless free particles in curved space-time using an adaptive $5^{\rm th}$ order Runge-Kutta algorithm. 
This is the first step toward developing a fully general relativistic 3D radiation transfer code for the Kerr space-time with arbitrary flows using characteristic methods and operator perturbation (Baron et al.\ 2009b). 
We derive the formalism for studying the strong lensing of the optical and X-ray emission by Kerr black holes in \textsection\ref{sec:basics}.
We find that the redshift effect caused by strong gravity plays an important role in computing the observed flux.
We do not consider the cosmological redshift, but to include the cosmic redshift is trivial:  In Eqs. (\ref{optical}) and (\ref{F_nu}), replace the strong lensing redshift factor $g$ by $g/(1+z_{\rm s})$, where $z_s$ is the cosmological redshift of the quasar.\footnote{Note that for  the optical emission, the $1+z_s$ factor is absorbed into the definition of the scale radius, \ie the $r_s$ is the emission size at frequency $\nu=(1+z_s)\nu_{\rm o}$. 
For X-ray emission, there will be a constant factor $(1+z_s)^{-(\Gamma+2)}$ in front of the integral in Eq.~(\ref{F_nu}).}  

\begin{figure}
	\epsscale{0.65}
	\plotone{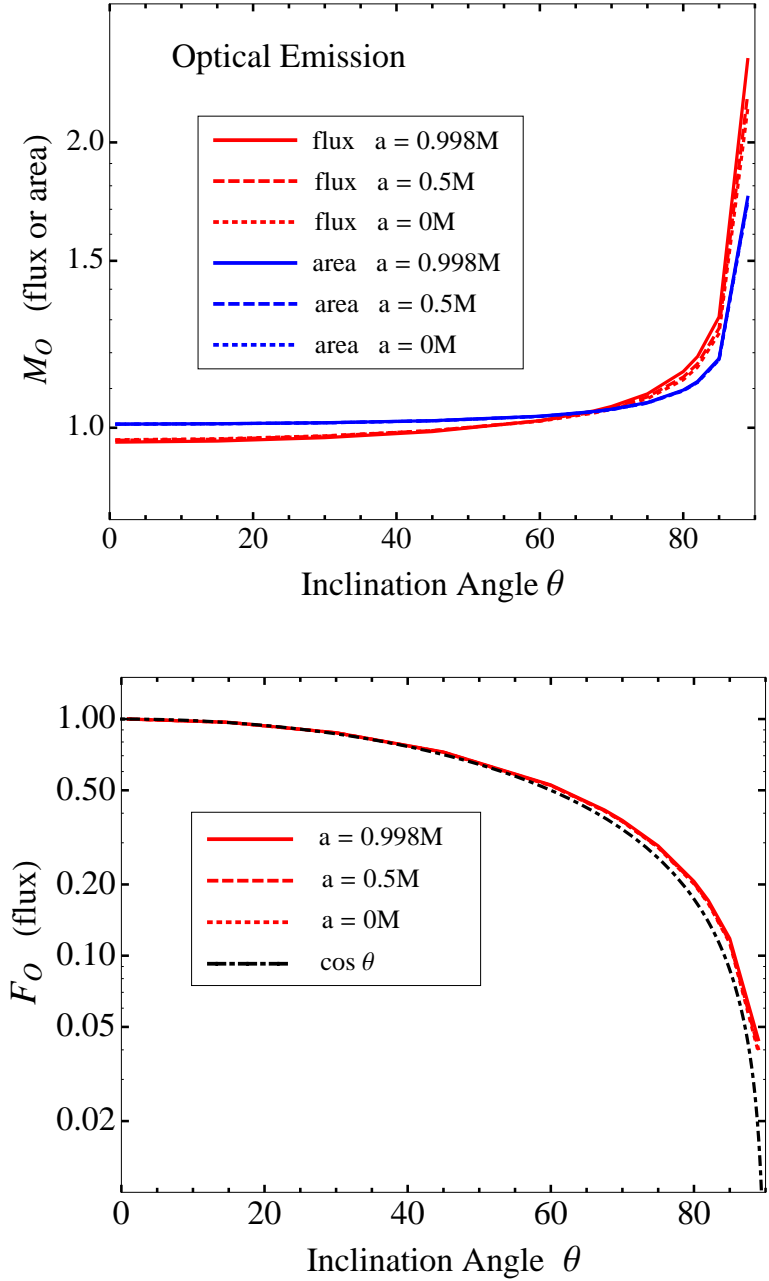}
	\caption{Inclination angle dependence of strong lensing of optical emission. 
	     In the top panel, we plot the lensing magnification $M_{\rm O}$ of image area (blue) and monochromatic  flux (red). 
	     In the bottom panel we plot the lensed flux including the projection effect (\ie the $\cos\theta$ factor)  (normalized to 1 when observed face on).
	     The effect of Kerr lensing is insignificant except for large inclination angles.
	     We assume the optical emission is Planckian with half light radius $r_{\rm half}=100\, r_g$ (scale radius  $r_{\rm s}=41\, r_g$), and disk size $r_{\rm disk} = 200\,r_g$.
            \label{fig:optical_mag}} 
\end{figure}

Using the sizes of the optical and X-ray emission regions measured by recent quasar microlensing observations (Kochanek 2004; Pooley et al.\ 2006; Morgan \etal 2008; Eigenbrod et al.\ 2008; Chartas \etal 2009; Dai \etal 2010; Blackburne \etal 2011; Blackburne et al.\ 2012),  we investigate the effects of the strong gravity field of the central Kerr black hole on the optical and X-ray continuum emission.
In particular, we calculate the correction to the X-ray-to-optical flux ratio caused by differential lensing distortions for the X-ray and optical emission. 
We assumed the standard thin disk model (Shakura \& Sunyaev 1973) for optical emission and find that the effect of the strong gravity is not very significant on the optical emission, since its emission size, $\sim$$100\,r_g,$ is about 10 times bigger than the X-ray emission (Figure~\ref{fig:optical_mag}).
The effect of strong gravity on optical emission is important only for large inclination angles.
We assume a simple power law and test three simple geometries for the X-ray emission:  an X-ray disk immediately above the accretion disk with inner cutoff at $r_{\rm ISCO}$ moving with the Keplerian flow, an X-ray ball surrounding the central black hole moving with Ruszkowshi \& Fabian (2000) velocity flow,  and an X-ray disk above the black hole with nonzero height moving with Ruszkowshi \& Fabian (2000) velocity flow.  
We find that the redshift effect is more important for the X-ray emission than for the optical emission (see e.g., Eqs (\ref{optical}) and (\ref{F_nu}), Figures \ref{fig:optical_mag} and \ref{fig:Xball_ratio}). 
Unlike the optical emission, the effect of Kerr strong lensing on the observed X-ray flux can be important even for moderate inclination angles (\ie for normal quasars). 
Among the three X-ray models tested in this paper, the low X-ray disk model shows  the largest variation in the X-ray-to-optical magnification ratio $R_{\rm X/O}$ (as large as a factor of 10 for normal quasars, and another factor of 10 for BAL+obscured quasars). 
Scatter by a factor of 10 (corresponds to $\Delta \alpha_{\rm ox}=0.384$) is still marginally consistent with current observations of $\alpha_{\rm ox}$ (see Eq.~(3) of Steffen et al.\ 2006) for normal quasars.
The strong inclination angle dependence makes $R_{\rm X/O}$ a random variable with nontrivial distribution, and the observed distribution of $\alpha_{\rm ox}$ is therefore modulated by that of $R_{\rm X/O}.$
In particular, $\alpha_{\rm ox}$ for a BAL can be greater than that of a normal quasar by $\sim$0.1--0.2 (see Table~\ref{tab:Gauss}).
Intrinsic absorption with column density $\sim$$10^{24}~\cmsq$ is needed in order to balance this boost of X-ray emission by Kerr strong lensing.
Comparing with current estimates of the BAL absorption column density  ($\sim 10^{22}$--$10^{24}~\cmsq$, e.g., Green et al.\ 2001; Gallagher et al.\ 2002), we find that it is important to include the Kerr strong gravity in order to correctly quantity the BAL absorption.
  
We note that for normal quasars, since the strong lensing distortion in the optical emission is not significant given the larger sizes of the optical emission region (of order $10^2\,r_g$), the current microlensing results need only minor modifications in order to incorporate the effect of strong gravity.
As can be seen from Figure~\ref{fig:optical_mag}, the strong lensing effect is significant for optical emission only when the inclination angle $\theta$ is high (observed nearly edge on).
Abolmasov \& Shakura (2012) studied three high amplification events from optical observations of SBS J1520+530, and Q~2237+0305 using straight line caustic crossing, and found that a model incorporating the strong lensing of the central black holes improves the fit; however, it requires high inclination angles ($\theta>70^\circ$) not typical for normal quasars. 
This is qualitatively consistent  with our results (see \textsection\ref{sec:optical}). 
    
For the X-ray emission, if the size of the X-ray emitting region is of order $10\, r_g$ (or even smaller, given the fact that most current microlensing observations give no lower limit on the X-ray emission sizes) and it is centered around the central black hole, then the X-ray emission will be strongly lensed by the central black hole before it is microlensed by stars in a lens galaxy much farther away.
Therefore, it might be important to add a strong lensing piece to the microlensing code which in principle should give more accurate constraints on the source sizes, in particular, the X-ray emission sizes. 
Furthermore, since Kerr strong lensing depends on important parameters such as the spin of the black hole, the inclination angle of the accretion disk, and the geometry of the corona, 
%(not to say the profile of the intensity, or the projected area along the line of sight), 
 it is possible to extract these valuable information by combining the Kerr and microlensing models.  
%the hard work in including strong lensing into the microlensing simulation will be rewarded by much more accurate and valuable information about AGN.  
A first attempt along this direction will be presented in a companion paper (Chen et al.\ 2012b).     

Of course, models such as a razor-thin X-ray disk above the accretion disk with an isotropic intensity profile may be too simplistic. %are not physically realistic.
More complicated and physically solid models should be tested in the future.  
But these simple models still show the importance of studying the effect of the strong gravity if the X-ray source is  indeed very small in size and very close to the central black hole. 
Although we have used the innermost stable circular orbit as a natural cutoff for the optical emission and the first X-ray model, it is possible to extend the emission region to within $r_{\rm ISCO}$ with the gas following the so-called plunging trajectories along geodesics (Krolik 1999; Agol \& Krolik 2000) and strong gravity plays an even more important role---the effects shown in this paper might be more significant.  
A radiation transfer code based on the ray-tracing code developed in this paper is being developed by the PHOENIX group and will be published in the future 
\footnote{PHOENIX is a generalized radiative transfer code that has been developed into 3D Framework in a series of  recent papers \citep[see][and references therein]{Peter2006,Chen2007,Baron09a,Baron09b,Baron12} that can handle radiative transfer in arbitrary metrics, with arbitrary velocity fields.}.
% gives a hint that the X-ray source of AGN can not be too close to the central black hole if it is very small in size as indicated by the microlensing observations, otherwise the strong lensing effect on the observed flux will have large inclination angle dependence as pointed out as early as in Cunningham (1975), which is in conflict with flux ratio observations.

We thank Ronald Kantowski for helpful discussions. 
We also thank the anonymous referee for the careful review of this work.  
BC and XD acknowledge support for this work provided by the National Aeronautics and Space Administration through Chandra Award Number GO0-11121B, GO1-12139B, GO2-13132A issued by the Chandra X-ray Observatory Center, which is operated by the Smithsonian Astrophysical Observatory for and on behalf of the National Aeronautics Space Administration under contract NAS8-03060.
BC and XD acknowledge support for program number HST-GO-11732.07-A  provided by NASA through a grant from the Space Telescope Science Institute, which is operated by the Association of Universities for Research in Astronomy, Incorporated.
BC and EB acknowledge NSF AST-0707704, and US DOE Grant DE-FG02-07ER41517 and support for program number HST-GO-12298.05-A provided by NASA through a grant from the Space Telescope Science Institute, which is operated by the Association of Universities for Research in Astronomy, Incorporated, under NASA contract NAS5-26555.

\appendix 
 
\section{Appendix}

We give the equations of motion (six $1^{\rm st}$ order ordinary equations) used in our the backward raytracing code. 
Inserting Eq.~(\ref{Hamilton}) into Eq.~(\ref{H_equation}), the equations of motion follow after straightforward but tedious algebra. For the 4 configuration-space variables $(t, r,\theta,\phi),$ we have  
\bea
\frac{dt}{d\lambda}&=&-\frac{p_t}{\alpha^2}- \frac{2Mar}{\rho^2\Delta}p_\phi\cr
\frac{dr}{d\lambda}&=&\frac{\Delta}{\rho^2}p_r\cr
\frac{d\theta}{d\lambda}&=&\frac{p_\theta}{\rho^2}\cr
\frac{d\phi}{d\lambda}&=&\left(\frac{\Delta-a^2\sin^2\theta}{\rho^2\Delta\sin^2\theta}\right)p_\phi -\frac{2Mar}{\rho^2\Delta}p_t,
\eea and for the two momentum-space variables $p_r,$ and $p_\theta$ ($p_t$ and $p_\phi$ are constants of the motion), 
\bea
\frac{dp_r}{d\lambda}&=&\frac{1}{2}\frac{\partial}{\partial r}\left(\frac{1}{\alpha^2}\right)p_t^2+ \frac{\partial}{\partial r}\left(\frac{2Mar}{\rho^2\Delta}\right)p_t p_\phi
-\frac{1}{2}\frac{\partial}{\partial r}\left(\frac{\Delta}{\rho^2}\right)p_r^2-\frac{1}{2}\frac{\partial}{\partial r}\left(\frac{1}{\rho^2}\right)p_\theta^2 \cr
&&-\frac{1}{2}\frac{\partial}{\partial r}\left(\frac{\Delta-a^2\sin^2\theta}{\rho^2\Delta\sin^2\theta} \right) p_\phi^2,
\eea and 
\bea
\frac{dp_\theta}{d\lambda}&=&\frac{1}{2}\frac{\partial}{\partial \theta}\left(\frac{1}{\alpha^2}\right)p_t^2+ \frac{\partial}{\partial \theta}\left(\frac{2Mar}{\rho^2\Delta}\right)p_t p_\phi
-\frac{1}{2}\frac{\partial}{\partial \theta}\left(\frac{\Delta}{\rho^2}\right)p_r^2-\frac{1}{2}\frac{\partial}{\partial \theta}\left(\frac{1}{\rho^2}\right)p_\theta^2 \cr
&&-\frac{1}{2}\frac{\partial}{\partial \theta}\left(\frac{\Delta-a^2\sin^2\theta}{\rho^2\Delta\sin^2\theta} \right) p_\phi^2,
\eea  where
\bea
\frac{\partial}{\partial r}\left(\frac{1}{\alpha^2}\right)&=&\left(\frac{2M}{\Delta \rho^2}\right)\left(\frac{a^4-r^4}{\Delta}-\frac{2r^2a^2\sin^2\theta}{\rho^2}\right)\cr
\frac{\partial }{\partial \theta}\left(\frac{1}{\alpha^2}\right)&=&\frac{4M a^2 r(a^2+r^2)}{\Delta \rho^4}\sin\theta\cos\theta,
\eea
\bea
\frac{\partial}{\partial r}\left(\frac{\Delta }{\rho^2}\right)&=&\frac{2}{\rho^2}\left(r-M-\frac{r\Delta}{\rho^2}\right)\cr
\frac{\partial}{\partial \theta}\left(\frac{\Delta }{\rho^2}\right)&=&\frac{2}{\rho^4}a^2\Delta \sin\theta\cos\theta,
\eea 
\bea
\frac{\partial}{\partial r}\left(\frac{1}{\rho^2}\right)&=&-2\frac{r}{\rho^4}\cr
\frac{\partial}{\partial \theta}\left(\frac{1}{\rho^2}\right)&=&2\frac{a^2}{\rho^4}\sin\theta\cos\theta,
\eea
\bea
\frac{\partial}{\partial r}\left(\frac{2Mar}{\rho^2\Delta}\right)&=& -\frac{\partial g^{03}}{\partial r} =\frac{2Ma\left[\rho^2(a^2-r^2)-2r^2\Delta\right]}{\rho^4\Delta^2}\cr
\frac{\partial}{\partial \theta}\left(\frac{2Mar}{\rho^2\Delta}\right)&=& -\frac{\partial g^{03}}{\partial\theta}  =\frac{4Mra^3\sin\theta\cos\theta}{\rho^4\Delta},
\eea and
\bea
\frac{\partial}{\partial r}\left(\frac{\Delta-a^2\sin^2\theta}{\rho^2\Delta\sin^2\theta} \right) &=& \frac{\partial g^{33}}{\partial r} =\frac{2a^2\sin^2\theta[r\Delta+(r-M)\rho^2]-2r\Delta^2}{\rho^4\Delta^2\sin^2\theta}\cr
\frac{\partial}{\partial \theta}\left(\frac{\Delta-a^2\sin^2\theta}{\rho^2\Delta\sin^2\theta} \right) &=&\frac{\partial g^{33}}{\partial \theta} =\frac{-2\cos\theta[\rho^2(\rho^2-2Mr)+2Mra^2\sin^2\theta]}{\rho^4\Delta\sin^3\theta}.
\eea

\newpage

%------------------ Table 1 --------------------------------

\begin{deluxetable}{llllllllllllll}
\tabletypesize{\scriptsize}
\tablecolumns{14}
\tablewidth{0pt}
\tablecaption{Strong Lensing  Flux Magnification for Optical and X-ray Emission. 
\label{tab:Flux_Mag}}
\tablehead{
\colhead{Model\tablenotemark{a}}&
\multicolumn{12}{c}{Inclination Angle (deg)} \\
\cline{2-13} \\
\colhead{}&
\colhead{ $1^\circ$}&
\colhead{ $15^\circ$}&
\colhead{ $30^\circ$}&
\colhead{ $45^\circ$} &
\colhead{ $60^\circ$}&
\colhead{ $67.5^\circ$}&
\colhead{ $70^\circ$} &
\colhead{$75^\circ$} &
\colhead{ $80^\circ$}&
\colhead{ $82^\circ$}&
\colhead{ $85^\circ$} &
\colhead{ $89^\circ$} &
\colhead{}
}
\startdata
optical emission\tablenotemark{b}  &    &     &   &     &    &      &        &     &  &   & & &     \\
\tableline
 ${\rm  a=0.998},                      r_{\rm disk}=200\,r_g         $ &  0.97      & 0.97  &   0.98     &  0.99  & 1.02       &  1.04        & 1.05    &  1.09    & 1.15  &  1.19  &  1.31  & 2.45    \\  
 ${\rm  a=0.5},\>\>\>\>             r_{\rm disk}=200\,r_g         $ &  0.97      & 0.97  &   0.98     &  0.99  & 1.02       &  1.04        & 1.05    &  1.08    & 1.13  &  1.17  &  1.27  & 2.25    \\  
  ${\rm  a=0}, \>\>\>\>\>\>\>\,   r_{\rm disk}=200\,r_g         $ &  0.97     & 0.97  &   0.98     &  0.99  & 1.02       &  1.04        & 1.05    &  1.07    & 1.12  &  1.16  &  1.26  & 2.17   \\   
\tableline
low X-ray disk  &    &     &   &     &    &      &        &     &  &   & & &     \\ 
\tableline
 ${\rm  a=0.998,                 n=3},\>  r_{\rm disk}= 5\,r_g         $ &  0.05     & 0.06  &   0.09     & 0.17   & 0.38       &  0.63        & 0.76    & 1.15     & 1.96  &  2.57  &  4.35  & 22.79     \\ 
 ${\rm  a=0.998,                 n=0},\> r_{\rm disk} = 5\,r_g         $ &  0.15     & 0.17  &   0.23     & 0.38   & 0.71       &  1.03        & 1.18    &  1.58     & 2.27  &  2.73  &  3.96  & 15.13    \\ 
 ${\rm  a=0.998,                 n=3},\> r_{\rm disk}= 10\,r_g        $ &  0.11     & 0.12  &   0.15     & 0.24   & 0.46       &  0.70        & 0.83    &  1.20     & 1.94  &  2.49  &  4.10  &  20.62   \\ 
 ${\rm  a=0.5,\>\>\>\>        n=3},\> r_{\rm disk}= 10\,r_g        $ &  0.34     & 0.36  &   0.44     & 0.62   & 0.94       &  1.20        & 1.30    &  1.59     & 2.03  &  2.31   &  3.04  & 9.54  \\
 ${\rm  a=0, \>\>\>\>\>\>\>  n=3},\> r_{\rm disk}= 10\,r_g       $ &  0.41     & 0.44  &  0.52     & 0.69   & 0.98       &  1.20        &  1.29    &  1.53    & 1.89  &  2.12    &  2.74 & 8.25  \\ 
 ${\rm  a=0.998,                   n=0},\> r_{\rm disk}= 10\,r_g      $ &  0.39     & 0.41  &   0.49     & 0.65   & 0.92       & 1.14         &  1.23   &  1.47    & 1.85   &  2.10    &  2.77  & 8.79   \\ 
 ${\rm  a=0.5,\>\>\>\>          n=0},\> r_{\rm disk}= 10\,r_g      $ &  0.42     &  0.44  &  0.53     &  0.69   & 0.97      & 1.19         &  1.27   &  1.50    &  1.85  & 2;.07    &  2.66    &  7.94     \\ 
 ${\rm  a=0,\>\>\>\>\>\>\>   n=0},\> r_{\rm disk}= 10\,r_g       $ &  0.44     & 0.47  &   0.55     & 0.72   & 0.99      & 1.19         &  1.28    &  1.49    & 1.83  & 2.04     &  2.61    &  7.71    \\ 
\tableline
X-ray ball   &    &     &   &     &    &       &        &     &  &   & & &       \\ 
\tableline
 ${  a=0.998,\, r_{\rm ball}= 3\,r_g}                                $ & 0.22   &  0.22  &  0.25  & 0.29  &  0.37      &  0.42      &   0.44   & 0.47  & 0.51 & 0.52  & 0.55   &  0.58   \\ 
 ${  a=0.5,  \>\>\>\>\>   r_{\rm ball} = 3\,r_g}                 $ & 0.14   &  0.15  &  0.17  & 0.20  &  0.27      &  0.32      &   0.35   & 0.40  & 0.46  & 0.50 & 0.54   &   0.62 \\ 
 ${  a=0, \>\>\>\>\>\>\>\>\>        r_{\rm ball} = 3\,r_g}   $ & 0.10  &  0.10   & 0.11   & 0.13   & 0.17      &  0.20      &    0.22  & 0.25  & 0.31 &  0.34 & 0.40   &  0.57    \\ 
  ${  a=0.998,\, r_{\rm ball}= 5\,r_g}                               $ & 0.45  &  0.46  &  0.49   & 0.53  &  0.59      &  0.62      &   0.64   & 0.66  & 0.69 &  0.70  & 0.71  &  0.74  \\ 
 ${  a=0.5,  \>\>\>\>\>   r_{\rm ball} = 5\,r_g}                 $ & 0.42  &  0.43  &  0.45   & 0.50  &  0.56      &  0.60      &   0.62   & 0.65  & 0.68  & 0.69  & 0.71   &  0.74  \\ 
 ${  a=0, \>\>\>\>\>\>\>\>\>        r_{\rm ball} = 5\,r_g}   $ & 0.39  &  0.40  & 0.43    & 0.48   & 0.55      &  0.60      &   0.62   & 0.65  & 0.69  & 0.70  & 0.73   &  0.76  \\  
 ${  a=0.998,\, r_{\rm ball}= 10\,r_g}                              $ & 0.72  &  0.73  &  0.73   & 0.75  &  0.78      &  0.80      &   0.81   & 0.82 & 0.84  & 0.84   & 0.85   &  0.86   \\ 
 ${  a=0.5,  \>\>\>\>\>   r_{\rm ball} = 10\,r_g}               $ & 0.71  &  0.72  &  0.72   & 0.74  &  0.77      &  0.79      &   0.80   & 0.82 & 0.83  & 0.84  & 0.85    &  0.86  \\ 
 ${  a=0, \>\>\>\>\>\>\>\>\>        r_{\rm ball} = 10\,r_g} $ &  0.71 &  0.71  & 0.72    & 0.74   & 0.77      &  0.79      &   0.80   & 0.82 & 0.83 &  0.84  & 0.85    &  0.87   \\   
\tableline
high X-ray disk  &    &     &   &     &    &       &        &     &  &   & & &      \\ 
\tableline
$  a=0.998, \> z_{\rm disk}= 3 \,   r_g   $                           & 0.50  & 0.52   & 0.56       & 0.64    & 0.78      & 0.90      &  0.96    &  1.11  & 1.40  &  1.61   & 2.22    &  8.67   \\ 
$  a=0.5,  \>\>\>\>\> \> z_{\rm disk}= 3 \,   r_g   $             & 0.49  & 0.50   & 0.55       & 0.63    & 0.77      & 0.89      &  0.94    &  1.10  & 1.40  &  1.61   &  2.24   &  8.84   \\ 
$   a=0.,  \>\>\>\>\>\>\>\> z_{\rm disk}= 3 \,   r_g   $         & 0.47  & 0.49   & 0.53       & 0.61    & 0.76      & 0.88       &  0.94    &  1.10   & 1.40 &  1.62  &  2.27   &  9.09   \\  
$   a=0.998, \> z_{\rm disk}= 5 \,   r_g   $                          & 0.58   & 0.59  & 0.62       & 0.68     & 0.79      & 0.89      & 0.94     & 1.07    & 1.34  &  1.54  & 2.13    &  8.35   \\ 
$  a=0.5,  \>\>\>\>\> \> z_{\rm disk}= 5 \,   r_g   $             & 0.57   & 0.58  & 0.61       & 0.67     & 0.78      & 0.88      &  0.93    & 1.07    & 1.34  &  1.54  &  2.13   &  8.44   \\ 
$  a=0.,  \>\>\>\>\>\>\>\> z_{\rm disk}= 5 \,   r_g   $          & 0.57  & 0.58   & 0.61       & 0.67     & 0.77      & 0.87      &  0.92    & 1.06    & 1.34  &  1.54  &  2.15   &  8.55   \\ 
$  a=0.998, \>\> z_{\rm disk}= 10\,  r_g   $                        & 0.71   & 0.71  & 0.73       & 0.76     & 0.84      & 0.91      &  0.95    & 1.06    & 1.27  &  1.44  & 1.93    & 7.14    \\ 
$  a=0.5,  \>\>\>\>\> \> z_{\rm disk}= 10 \,   r_g   $           & 0.70   &  0.71  & 0.73      & 0.76     & 0.83      & 0.91      &  0.94    &  1.05   & 1.27  &  1.43  &  1.92   & 7.15    \\ 
$   a=0.,  \>\>\>\>\>\>\>\> z_{\rm disk}= 10 \,   r_g   $       & 0.70   & 0.71   & 0.72      & 0.76    & 0.83      & 0.90      &  0.94     &  1.05   & 1.27  &  1.43  &  1.93   & 7.17   \\  
\enddata
\tablenotetext{a}{To obtain the inclination-dependent flux profile (see Figures~\ref{fig:optical_mag}, \ref{fig:Xray_flux}, \ref{fig:Xball_ratio}, \ref{fig:Xover_ratio}), the geometrical projection effect should be included, \ie a $\cos\theta$ factor for optical emission and the two X-ray disk models, or a $(1+\cos\theta)/2$ factor for the X-ray ball model.  }
\tablenotetext{b}{ $M_{\rm O}=0.2+0.8M_{<200\,r_g}$, where $M_{<200\,r_g}$ is the strong lensing magnification of the optical flux within $r_{\rm disk}=200\,r_g.$ }
\end{deluxetable}

%-------------- Table 2--------------

\begin{deluxetable}{clllllllllllll}
\tabletypesize{\scriptsize}
\tablecolumns{14}
\tablewidth{0pt}
\tablecaption{$\alpha_{\rm ox}$  Distributions After Kerr Strong Lensing.
 %Consequently, $0^\circ < \theta^{\rm Normal}\le 58.7^\circ <\theta^{\rm BAL}\le 78.5^\circ.$
\label{tab:Gauss}}
\tablehead{
\colhead{Model\tablenotemark{a}}&
\multicolumn{3}{c}{Normal QSO\tablenotemark{b}} & \colhead{} & \multicolumn{3}{c}{BALQSO\tablenotemark{b}} &  \colhead{} &  \multicolumn{3}{c}{Normal+BAL QSO\tablenotemark{b}}  & \colhead{}  & \colhead{$\Delta\langle\alpha_{\rm ox}\rangle$\tablenotemark{c}} \\
\cline{2-4}\cline{6-8}\cline{10-12} \\
\colhead{}&
\colhead{ $\langle\alpha_{\rm ox}\rangle$\tablenotemark{c}}&
\colhead{ $\rm \sigma$}&
\colhead{ $\cal S$}&
\colhead{} &
\colhead{ $\langle\alpha_{\rm ox}\rangle$}&
\colhead{ $\rm \sigma$}&
\colhead{ $\cal S$} &
\colhead{} &
\colhead{ $\langle\alpha_{\rm ox}\rangle$}&
\colhead{ $\rm \sigma$}&
\colhead{ $\cal S$} &
\colhead{} &
\colhead{}
}
\startdata
low X-ray disk  &    &     &   &     &    &      &        &     &  &   & & &     \\ 
\tableline
 ${\rm  a=0.998,            n=3}         $ &  -1.76     & 0.211  &   0.003     &    & -1.55       &  0.208        & 0.004    &      & -1.67  &  0.234  &  0.03  &  &  0.21   \\ 
 ${\rm  a=0.5,\>\>\>\>   n=3}         $ &  -1.60     & 0.205  &   0.001     &    & -1.47       &  0.202        & 0.000    &      & -1.55 &  0.214   &  0.002  &  &  0.13  \\
 ${\rm  a=0, \>\>\>\>\>\>\>  n=3}  $ &  -1.58     & 0.204  &    0.001    &    & -1.47       &  0.202       &  0.000    &     & -1.53 &  0.210    &  0.001 &  &  0.11 \\ 
 ${\rm  a=0.998,           n=0}          $ &  -1.59     & 0.204  &   0.001     &    & -1.48      & 0.202         &  0.001  &     & -1.54 &  0.210    &  0.002  &  &  0.11  \\ 
 ${\rm  a=0.5,\>\>\>\> n=0}           $ &  -1.57     &  0.203  &  0.001     &    & -1.47      & 0.201         &  0.000   &     &  -1.53 & 0.209  &  0.001  &  & 0.10    \\ 
 ${\rm  a=0,\>\>\>\>\>\>\> n=0}    $ &  -1.57      & 0.203  &   0.001     &    & -1.47      & 0.201         &  0.000    &     & -1.53  & 0.208  &  0.001  &  &  0.10    \\ 
\tableline
X-ray ball\tablenotemark{d}   &    &     &   &     &    &       &        &     &  &   & & &       \\ 
\tableline
 ${  a=0.998,\, r_{\rm ball}= 3\,r_g}                                $ &  -1.68  &  0.204  &  0.000  &   &  -1.54      &  0.205      &   0.003   &  & -1.63  & 0.216   & 0.014 &   &    0.14  \\ 
 ${  a=0.5,  \>\>\>\>\>   r_{\rm ball} = 3\,r_g}                 $ &  -1.75  &  0.205  &  0.000  &   &  -1.58      &  0.207      &   0.006   &  & -1.68  & 0.222   & 0.029 &   &    0.17  \\ 
 ${  a=0, \>\>\>\>\>\>\>\>\>        r_{\rm ball} = 3\,r_g}   $ &  -1.82  &  0.204  &  0.001  &    & -1.66       &  0.208     &    0.007   &  & -1.75 &  0.221  & 0.035  &   &    0.16  \\ 
  ${  a=0.998,\, r_{\rm ball}= 5\,r_g}                               $ &  -1.58  &  0.202  &  0.001  &   &  -1.47      &  0.203      &   0.002   &  & -1.54  & 0.209   & 0.011 &  &     0.11  \\ 
 ${  a=0.5,  \>\>\>\>\>   r_{\rm ball} = 5\,r_g}                 $ &  -1.59  &  0.202  &  0.001  &   &  -1.48       &  0.204     &   0.002   &  & -1.55  & 0.210   & 0.013 &   &    0.11  \\ 
 ${  a=0, \>\>\>\>\>\>\>\>\>        r_{\rm ball} = 5\,r_g}   $ &  -1.60  &  0.203   & 0.001  &    & -1.48       &  0.204     &   0.002   &  & -1.55  & 0.212   & 0.015  &  &   0.12  \\  
\tableline
high X-ray disk  &    &     &   &     &    &       &        &     &  &   & & &      \\ 
\tableline
 $  a=0.998, \> z= 3 \,   r_g   $ & -1.58  & 0.201   & 0.000        &    & -1.52      & 0.201      &  0.000    &     & -1.55  & 0.203  &  0.001   &  &  0.06  \\ 
 $  a=0.998, \> z= 5 \,   r_g   $ & -1.57  & 0.201   & 0.000        &    & -1.52      & 0.201      &  0.000    &     & -1.55  & 0.202  &  0.001   &  &  0.05   \\ 
 $  a=0.998, \> z= 10\,  r_g   $ & -1.54  & 0.200   & 0.000        &    & -1.52      & 0.200      &  0.000    &    & -1.53  & 0.201   &  0.000   &  &  0.02    \\ 
\enddata
\tablenotetext{a}{For the X-ray ball model, we test only for case $r_{\rm ball}= 3\, r_g,$ and $5\,r_g$ considering the relative flatness of $R_{\rm X/O}(\theta)$ for $r_{\rm ball}= 10\,r_g$ (Figure~\ref{fig:Xball_ratio}).
                               For the high X-ray disk model, we consider only the $a=0.998$ case because of the similarity of the $R_{\rm X/O}(\theta)$ curves between different spins (Figure~\ref{fig:Xover_ratio}). }
\tablenotetext{b}{ We assume that obscured (Type II) quasars are about  $20\%$ of quasar population and BALs count for 40\% of Type I quasars.
			    The intrinsic distribution of $\alpha_{\rm ox}$ is assumed to be Gaussian with $\mu=-1.5$ and $\sigma=0.2.$
			    $\langle\alpha_{\rm ox}\rangle$, $\sigma,$ and $\cal S $ are respectively the mean, standard deviation, and skewness ($\mu_3/\sigma^3$) of  $\alpha_{\rm ox}$ distribution.    }                               
%\tablenotetext{c}{$\langle\alpha_{\rm ox}\rangle$, $\sigma_{\alpha_{\rm ox}},$ and $\cal S $ are the mean, standard deviation, and skewness ($\mu_3/\sigma^3$) of  $\alpha_{\rm ox}$. }
\tablenotetext{c}{$\Delta\langle\alpha_{\rm ox}\rangle\equiv \langle\alpha_{\rm ox}\rangle^{\rm BAL}-\langle\alpha_{\rm ox}\rangle^{\rm Normal}, $ the difference in mean of $\alpha_{\rm ox}$ between BALQSOs and normal quasars.  }
\tablenotetext{d}{There exists extra scatter of $\alpha_{\rm ox}$ (besides strong gravity) due to projection effect of different geometries of X-ray (half sphere) and optical emission (thin disk). 
                             We have included this effect into our Monte-Carlo simulation.   }
\end{deluxetable}

\newpage

\newpage

\newpage

\newpage

\newpage

\clearpage

%---------------------------------------------------------------------------
\end{document}